\documentclass[11pt,a4paper]{article}
\usepackage{jheppub}
\usepackage{lipsum}
\usepackage{braket}
\usepackage{tensor}
\usepackage{dsfont}
\usepackage{caption}
\usepackage[dvipsnames*,svgnames]{xcolor}
\usepackage{subcaption}
\usepackage{tikz}
\usepackage{pgfplots}
\usetikzlibrary{datavisualization}
\usetikzlibrary{datavisualization.formats.functions}
\usetikzlibrary{decorations.pathreplacing,calligraphy}
\usetikzlibrary{decorations.pathreplacing}
\usetikzlibrary{decorations.pathmorphing, patterns,shapes}
\usepackage{bm}

\newmuskip\pFqmuskip
\newcommand*\pFq[6][8]{%
  \begingroup 
  \pFqmuskip=#1mu\relax
  \mathchardef\normalcomma=\mathcode`,
  \mathcode`\,=\string"8000
  \begingroup\lccode`\~=`\,
  \lowercase{\endgroup\let~}\pFqcomma
  {}_{#2}F_{#3}{\left[\left.\genfrac..{0pt}{}{#4}{#5}\right| #6\right]}%
  \endgroup
}
\newcommand{\pFqcomma}{{\normalcomma}\mskip\pFqmuskip}

\newcommand{\EM}[1]{{\color{blue}  EM: #1}} 
\newcommand{\AM}[1]{{\color{orange}  AM: #1}} 

\newcommand{\co}{\mathcal{O}}

\newcommand{\phu}{\varphi}
\newcommand{\op}[1]{\boldsymbol{#1}}
\newcommand{\de}{\mathrm d}

\title{Sum rules \& Tauberian theorems at finite temperature}

\author[a]{Enrico Marchetto,}
\author[b]{Alessio Miscioscia}
\author[b]{and Elli Pomoni}
\affiliation[a]{Mathematical Institute, University of Oxford, \\
	Andrew Wiles Building, Woodstock Road, Oxford, OX2 6GG, UK}
\affiliation[b]{Deutsches Elektronen-Synchrotron DESY, Notkestr. 85, 22607 Hamburg, Germany}
\emailAdd{enrico.marchetto@maths.ox.ac.uk}
\emailAdd{alessio.miscioscia@desy.de}
\emailAdd{elli.pomoni@desy.de}
\preprint{DESY-23-224}
\abstract{
We study CFTs at finite temperature and derive explicit sum rules for one-point functions of operators by imposing the KMS condition and we explicitly estimate one-point functions for light operators. 
Turning to heavy operators we employ Tauberian theorems and
 compute the asymptotic OPE density for heavy operators,
from which we extract the leading terms of the OPE coefficients associated with heavy operators. Furthermore, we approximate and establish bounds for the two-point functions.
}
\begin{document}
	\maketitle
	\flushbottom
\section{Introduction}

Thermal effects in Conformal Field Theories (CFTs) can be studied following the thermal bootstrap program.
This approach, initially introduced in \cite{Iliesiu:2018fao} and further developed in \cite{Iliesiu:2018zlz, Petkou:2018ynm,Alday:2020eua,Benjamin:2023qsc, Marchetto:2023fcw}, revolves around the concept that thermal effects can be incorporated by placing the CFT on the thermal manifold $S_\beta^1\times \mathbb R^{d-1}$, where $\beta = 1/T$ is the inverse of the temperature, and identifying the time direction as the direction along the circle. The Operator Product Expansion (OPE) and its constituents, namely the (zero temperature) CFT data, remain unaltered by thermal effects. However, the OPE converges only when the distance between two operators is less than $\beta$. The periodicity along the thermal circle of the thermal correlation functions is referred to as the Kubo-Martin-Schwinger (KMS) condition, proposed in \cite{El-Showk:2011yvt} as the equivalent of the crossing equation.
\newline 
In this work, we extend our investigation along this research line by deriving new implications of the KMS condition and delving deeper into established constraints. As anticipated, it has been emphasized that thermal one-point functions can be constrained by examining the OPE around the KMS symmetric point which corresponds to setting the time coordinate equal to $\beta/2$ \cite{El-Showk:2011yvt, Iliesiu:2018fao}. We explicitly formulate sum rules, comprising an infinite set of equations, for thermal one-point functions. What is more, we investigate these consistency conditions under the assumption of setting the spatial coordinates to zero, thus disregarding the spin degeneracy of the operators.
In fact, by setting the spatial coordinates to zero, the OPE of the two-point function relies only on the conformal dimensions of the operators, with the coefficients in front being weighted sums of OPE coefficients for operators with identical conformal dimensions. We conduct explicit numerical validations of the sum rules, focusing on straightforward cases such as (generalized) free theories or two-dimensional systems. The sum rules can be applied to estimate one-point functions of light operators. Furthermore, we quantify the error associated with the truncation of the sum rule up to some conformal dimension.
\newline 
Inspired by the conformal case at zero temperature \cite{Qiao:2017xif}, we calculate the large dimension's asymptotic of the thermal OPE density when the spatial coordinates are fixed to zero. Essentially, this density represents the states of the CFT weighted by a sum of thermal OPE coefficients. The result is derived by invoking Tauberian theorems and the KMS condition, under the assumption that the density is bounded from below. The lack of positivity of thermal coefficients makes the latter assumption non-trivial in the finite temperature case. Nonetheless, we demonstrate that OPE coefficients for heavy operators share the same sign. This implies that the density is either bounded from above or from below.\footnote{By multiplying by a sign, it can always be adjusted to the case where the density is bounded from below.} This conclusion is reached by considering the inversion formula in conjunction with KMS.
The asymptotic density of states enables us to extract the leading term of OPE coefficients associated with heavy operators, and approximate and rigorously bound the two-point function (when computed at zero spatial coordinates). We validate these results through explicit testing in various scenarios, including (generalized) free theories, two-dimensional thermal CFTs, and the three-dimensional $\mathrm O(N)$ model at large $N$.
\newline The paper is organized as follows: \begin{itemize}
    \item [$\star$] In Section \ref{sec:KMSTCFT} we derive the sum rules from KMS and we specify them to the zero spatial coordinate case. After performing several numerical checks we estimate light operators' one-point functions;
    \item [$\star$] In Section \ref{sec:tauberianOPE} we introduce the thermal OPE density and we derive its asymptotic for high dimensions. From the latter asymptotic we derive the leading term of OPE coefficients for heavy operators and we bound the two-point functions for values of the time $\tau$ close to zero or $\beta$. Explicit checks are provided in (generalized) free theories, three-dimensional $\mathrm O(N)$ at large $N$, and two-dimensional cases. Finally, we use the Tauberian theorems to estimate the error associated with the truncation of the sum rules computed in the previous Section;
\end{itemize}
We conclude with a discussion in Section \ref{eq: discussion}.
The paper also contains four appendices with a number of detailed computations and explanations. In Appendix \ref{appendix:Derivation} the reader can find all the detailed computations leading to the sum rules for thermal one-point functions. In \ref{appendix:OPEdensityProof} there is a short review of Tauberian theorems with one of the possible way to prove them: we choose to present Karamata's argument. Finally in Appendices \ref{appendix:boundedness} and \ref{appendix:DetailsOn2ptbounds} there are the details on the boundedness of the OPE coefficients and some computations, useful in order to bound the thermal two-point functions.

\section{Sum rules from KMS}\label{sec:KMSTCFT}
It has been known for a long time that studying thermal effects in quantum field theory is equivalent to replacing the ambient, flat manifold $\mathbb{R}^{d}$ with the thermal manifold $ \mathbb R^{d-1}\times S_\beta^1$, where the time direction has been compactified on a circle of length $\beta= 1/T$. Periodic boundary conditions must be imposed on bosonic fundamental fields, and anti-periodic on the fermionic ones. A simple condition we should require for two-point functions between scalars (and for any correlation function in general) is \emph{periodicity} along the thermal circle, in formula \footnote{We already used translational invariance and rotations between spatial coordinates to conclude that the two-point function only depends on $\tau$ and $r = |\vec x|$ \cite{Iliesiu:2018fao, Marchetto:2023fcw}.} 
\begin{equation} \label{eq: KMS}
    \langle \phi(\tau,r) \phi(0,0)\rangle_\beta = \langle \phi(\tau+\beta,r) \phi(0,0)\rangle_\beta  \ ,
\end{equation}
This is known in the literature as the KMS condition \cite{Kubo:1957mj,Martin:1959jp}. The key idea is to combine it with the OPE, which reads \cite{Katz:2014rla,Iliesiu:2018fao}
\begin{equation}\label{eq:OPE}
    \langle \phi(\tau,r)\phi(0,0)\rangle_\beta =  \sum_{\mathcal O \in \phi \times \phi}\frac{a_{\mathcal O}}{\beta^{\Delta}} |\tau^2+r^2|^{\frac{\Delta-2\Delta_{\phi}}{2}} C_{J}^{(\nu)}\left(\frac{\tau}{\sqrt{r^2+\tau^2}}\right) \ ,
\end{equation} 
where $\Delta$ and $J$ are the conformal dimension and the spin of the operator $\mathcal{O} \in \phi \times \phi$,  $C_J^{(\nu)}$ are the Gegenbauer polynomials and \begin{equation}
    a_{\mathcal O} = \frac{f_{\phi\phi\mathcal O}b_{\mathcal O}}{c_{\mathcal O}} \frac{J!}{2^J (\nu)_J} \ , \qquad (\nu)_J = \frac{\Gamma(\nu+J)}{\Gamma(\nu)} \ , \qquad \nu=\frac{d-2}{2} \ , \label{eq: def a}
\end{equation}
where $b_{\mathcal O}$ is the thermal one-point function coefficients, $f_{\phi\phi\mathcal O}$ the zero-temperature structure constant, and $c_{\mathcal O}$ the normalization of the zero-temperature two-point function of two identical operators $\mathcal O$. The OPE \eqref{eq:OPE} converges only in a radius $\tau^2+r^2< \beta^2$ on the thermal manifold. Knowing this, it is convenient to combine the KMS condition with the parity (or time reversal) transformation $\tau \to -\tau$. This is to ensure that the OPE expansion around $\tau = 0$ is convergent and well-defined on both sides of the equation. Indeed 
\begin{equation}\label{eq:GeneralDContraint}
       \langle \phi(\beta/2+\tau,r)\phi(0)\rangle_\beta = \langle \phi(\beta/2-\tau,r)\phi(0)\rangle_\beta  \ .
   \end{equation}
   \noindent The equation \eqref{eq:GeneralDContraint} had already been identified by El-Showk and Papadodimas \cite{El-Showk:2011yvt} as the crossing equation analog for thermal CFTs. In this Section, we explicitly explore this idea by extracting \emph{explicit sum rules} out of the condition \eqref{eq:GeneralDContraint}, making the analogy proposed by the aforementioned authors closer. The details of the derivation of the sum rules are explicitly given in Appendix \ref{appendix:Derivation}. Schematically, the idea is the following: we expand both the left and the right side of  \eqref{eq:GeneralDContraint} by using the OPE, then we further expand the result in powers of $\tau$ and $r$. This can be achieved by using the definition of the Gegenbauer polynomials \eqref{eq: gegdef} and the binomial theorem, applied twice. It is easy to see from equation \eqref{eq:GeneralDContraint} that the even powers of the $\tau$ variable identically cancel,  while the terms proportional to odd powers have to be set equal to zero and return non-trivial equations. Imposing this order by order in the $r$ variable, we obtain explicit sum rules for the combination $b_{\mathcal O} f_{\mathcal O \phi\phi}$  (see Appendix \ref{appendix:Derivation} for the detail of the derivation)
   \begin{equation}\label{eq:sumrule}
    \boxed{\sum_{\mathcal O \in \phi \times \phi} b_{\mathcal O} f_{\mathcal O \phi \phi} \,  F_{\ell, n}(h,J) = 0 \ , \qquad  n \in \mathbb N \ , \ell \in 2 \mathbb{N}+1} \ ,
  \end{equation}
  where we introduced the twist
  \begin{equation}
      h=\Delta-J  \ ,\label{eq: twist}
  \end{equation}
  we defined the function
  \begin{equation}
      F_{\ell, n}(h,J)=\frac{1}{2^{h+J}} \binom{\frac{h-2 \Delta_{\phi}}{2}}{n} \binom{h+J-2 \Delta_{\phi} -2 n }{\ell} \, {}_{3}F_{2}{\left[\left.\genfrac..{0pt}{}{\frac{1-J}{2},-\frac{J}{2}, \frac{h}{2}-\Delta_{\phi} +1}{\frac{h}{2}-\Delta_{\phi}-n +1,-J-\nu +1}\right| 1\right]} \ , \label{eq: sumrules}
  \end{equation}
  and we set the normalization $c_{\mathcal O} = 1$.
  As expected, the function $F_{\ell, n}(h,J)$ depends both on the conformal dimensions $\Delta$ (via the $h$ twist variables) and the spins $J$ of the operators appearing in the OPE; moreover, it also depends on two positive, integer numbers $\ell$ and $n$. Adopting a point of view closer to the zero-temperature analytical (and numerical) conformal bootstrap (see \cite{Fitzpatrick:2012yx,Komargodski:2012ek,Alday:2013cwa,Alday:2013opa,Bissi:2022mrs,Hartman:2022zik,Poland:2018epd} and references therein), the above formula corresponds to setting to zero the $(2\ell+1)$-th derivative in $\tau$ and the $n$-th derivative in $r$ of $f(\tau, r)$ when computing in $\tau = \beta/2$ and $r = 0$ \cite{Iliesiu:2018fao}
  \begin{equation}
      \left. \frac{\partial^{2 \ell+1}}{\partial \tau^{2 \ell+1}}\frac{\partial^{n}}{\partial r^{n}} \langle \phi(\tau,r) \phi(0,0)\rangle_\beta \right |_{\tau = \frac{\beta}{2}, r = 0} = 0 \ .
  \end{equation}
  In this way, it is clear that the sum rules in equation \eqref{eq:sumrule} are the incarnation of an expansion around the fixed point of the relation \eqref{eq:GeneralDContraint}. In the following, we will check the sum rules in the four-dimensional free scalar theory. Then, we will set $r=0$ and study setups where the two operators in the correlation function lie on the same thermal circle. The simplest way to use equations \eqref{eq:sumrule} is to truncate the OPE. This is analogous to the \textit{Gliozzi method} in conformal bootstrap, proposed in \cite{Gliozzi:2013ysa}. It is not in general guaranteed that just a simple truncation of the infinite sum over operators will give reliable results: we will discuss this point in more detail, in the case of zero spatial coordinates, in Section \ref{ssec: larggap}. To understand if the sum rules in \eqref{eq: sumrules} (when $n = 0$) can be truncated we need to gather quantitative information about the tail of heavy operators. In Section \ref{sec:tauberianOPE} we will prove a general result about those operators and therefore we will quantify the error made by a simple truncation of the OPE (see Section \ref{sec:error}).
    \paragraph{A comment on free scalar theory}
    Let us comment on the simplest example, with $\phi$ being a fundamental free scalar field. In this particular scenario, the upper argument of the first binomial in the equation \eqref{eq: sumrules} is fixed to zero. Indeed, apart from the case of the identity operator $\mathds{1}$, the conformal dimensions of the operators $\mathcal{O}\in \phi \times \phi$ are given by $\Delta = 2\Delta_\phi+J$, which leads to $h - 2\Delta_\phi=0$. This implies that all the equations for $n \neq 0$ are trivially satisfied. This was to be expected since the operators $\mathcal{O}\in \phi \times \phi$ can be labeled only by their conformal dimensions, which are in a bijection with the spins. Therefore it is natural to think that setting the spatial coordinates to zero from the beginning (which is equivalent to set $n = 0$) is enough to solve for the OPE coefficients. This theory can be explicitly solved \cite{Iliesiu:2018fao, Marchetto:2023fcw} and the thermal OPE coefficients explicitly computed. Since we only have to discuss the sum rules for $n = 0$, we postpone this check to Section \ref{ssec: zero} where a more general discussion on the correlator with $r=0$ is considered. We anticipate here that the sum rules can be numerically checked and we observe that the bigger $\ell$ is, the bigger the dimensions of the operators considered has to be for the sum rule to give reliable results (see Fig. \ref{fig:D1GFF}).
  \subsection{Reduction to zero spatial coordinates} \label{ssec: zero}
\begin{figure}[h!]
\begin{subfigure}[t]{.48\textwidth}
  \centering
  \includegraphics[width=\textwidth]{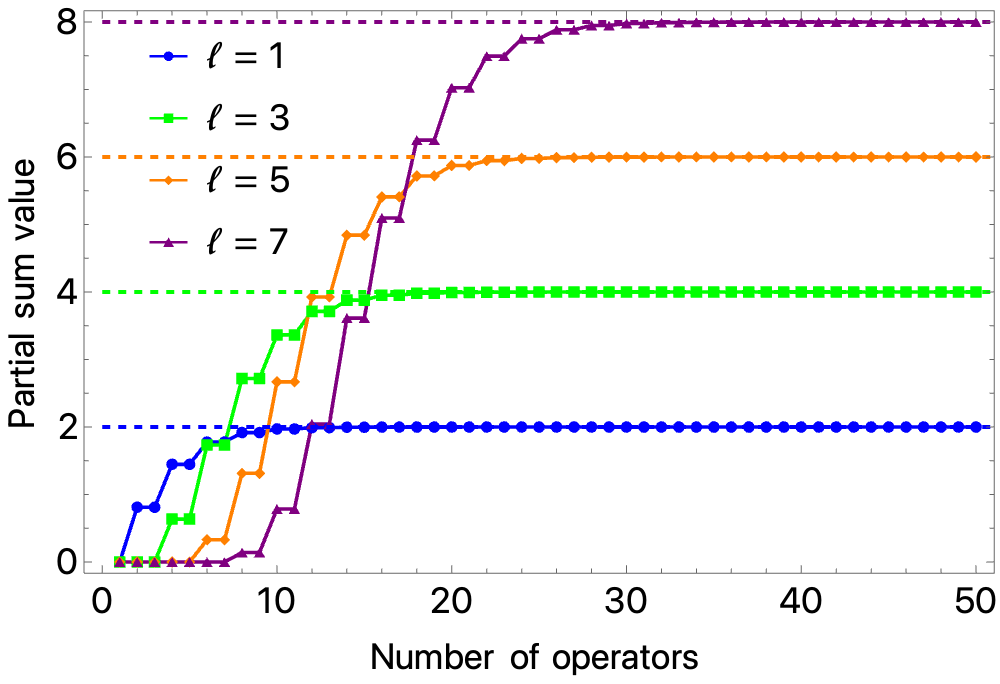}
  \caption{}
  \label{fig:D1GFF}
\end{subfigure}%
\hfill
\begin{subfigure}[t]{.49\textwidth}
  \centering
  \includegraphics[width=\linewidth]{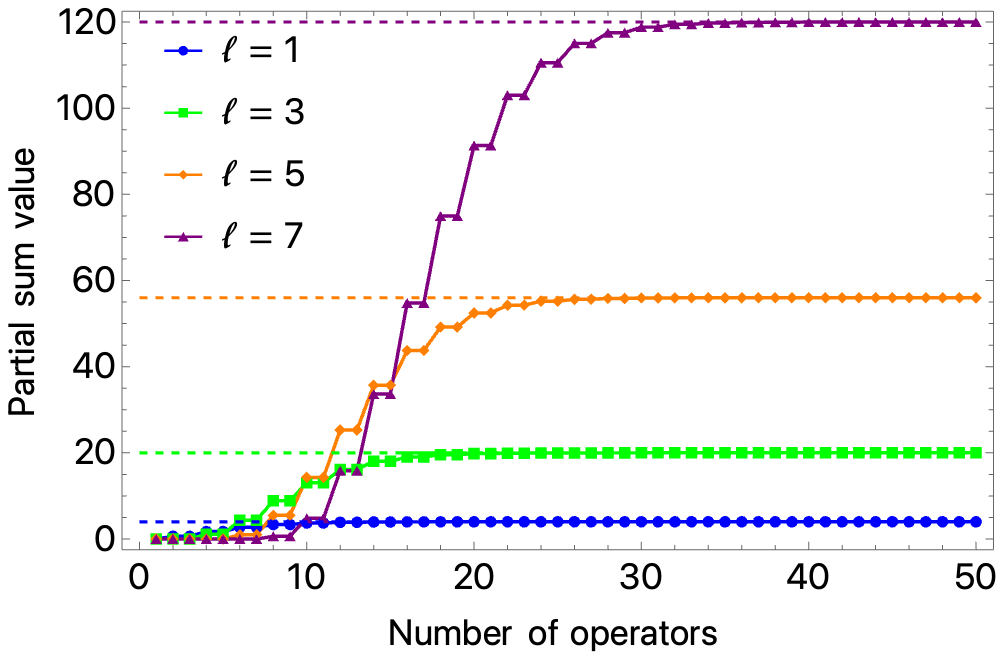}
  \caption{}
  \label{fig:D3GFF}
\end{subfigure}
\label{fig:GFFCheck}
\caption{\emph{Comparing the left (straight, dashed lines) and right (continuous lines) sides of equation \eqref{eq:BootstrapFor1D} for different values of $\ell$. The bigger $\ell$ is, the more operators need to be summed for the right-hand side to converge to the correct value, namely the left-hand side. The \textbf{left panel (a)} corresponds to a numerical test for a \underline{GFF with $\Delta_\phi = 1$}, or equivalently for a four-dimensional scalar free theory. The \textbf{right panel (b)} corresponds to a numerical test for a \underline{GFF with $\Delta_\phi = 2$}, or equivalently for a six-dimensional scalar free theory. In both cases, it is convenient to multiply both sides of the equation \eqref{eq:BootstrapFor1D} by a factor $1/\ell!$.}}
\end{figure}

We consider in this work only a specific subset of the equations coming from \eqref{eq:sumrule}, namely the one with $n = 0$. These equations correspond to the one-dimensional case in which we set the spatial coordinates of the two-point function equal to zero. In this case, it is possible to reduce the general sum rules \eqref{eq:sumrule} to a simpler form 
\begin{equation}\label{eq:BootstrapFor1D}
    \frac{\Gamma \left(2 \Delta_\phi+ \ell \right)}{\Gamma \left(2 \Delta_\phi\right)} = \sum_{\Delta} \frac{a_{\Delta}}{2^{\Delta}} \frac{\Gamma\left(\Delta-2\Delta_\phi+1\right)}{\Gamma \left(\Delta-2\Delta_\phi-\ell+1\right)} \ ,
\end{equation}
where $\ell \in 2\mathbb N+1$. See Appendix \ref{appendix:Derivation} for the derivation of this formula from the more general result \eqref{eq: sumrules}.\footnote{Alternatively one can directly start with the correlator at zero spatial coordinates and use the binomial theorem, as in the proof of equation \eqref{eq:sumrule} in Appendix \ref{appendix:Derivation}.} The sum above is performed over all the possible conformal dimensions appearing in the OPE between $\phi \times \phi$ and, due to the reduction, we are not sensitive to the spins of the operators anymore. Indeed when a two-point function of a $d$-dimensional theory is restricted to zero spatial coordinates the thermal OPE takes the form 
\begin{equation}\label{eq:twopintOPEZeror}
    \langle \phi(\tau,0) \phi(0,0)\rangle_\beta = \sum_{\Delta} \frac{a_{\Delta}}{\beta^{\Delta}} \tau^{\Delta-2\Delta_\phi} \ , \qquad a_{\Delta} \equiv \sum_{\mathcal{O} \in \phi \times \phi}^{\Delta \text{ fixed}}  a_{\mathcal O}  \, C_J^{(\nu)}(1) \ ,
\end{equation}
where $a_{\mathcal O}$ is given in \eqref{eq: def a}
and
the last sum is performed over operators in the OPE $\phi \times \phi$ of dimension $\Delta$ (but different spin $J$).
These equations are, in principle, non-trivial constraints for the OPE coefficients and therefore for one-point functions of operators in the OPE $\phi\times \phi$. In deriving equation \eqref{eq:BootstrapFor1D} we used the fact that in the OPE the identity operator $\mathds{1}$ appears. The conformal dimension of the identity is zero (and so its spin), and its one-point function and structure constant can be set to $b_{\mathds{1}} = f_{\phi \phi \mathds{1}} = 1$. We can then isolate the contribution of the identity, the left side of equation \eqref{eq:BootstrapFor1D}. This information is important to define a normalization, otherwise the coefficients $a_{\Delta}$ could only be defined up to an overall rescaling (this is clear from equation \eqref{eq:sumrule}). We provide now some examples and interesting consequences of equation \eqref{eq:BootstrapFor1D}. Let us comment that equations \eqref{eq:sumrule} and \eqref{eq:BootstrapFor1D} can be straightforwardly generalized to the case of unequal external operators. The structure of the equations is exactly the same but the identity will not be included in the OPE.
\paragraph{First test: generalized free fields in different dimensions}\label{GFF1d}
Equation \eqref{eq:BootstrapFor1D} is the only set of equations for thermal CFTs correlators restricted to zero spatial coordinates. Among all possible thermal CFTs, some can be solved with other methods. One class of such theories is represented by the generalized free fields (GFF) defined by the OPE 
\begin{equation}
    \phi \times \phi = \mathds{1} + [\phi \phi]_{p, q}\ ,
\end{equation}
where the conformal dimensions of the operators in the OPE are $\Delta= 2 \Delta_\phi+p+2 q$, $p, q \in \mathbb N$ (up to the identity $\mathds{1}$, whose dimension is zero). These theories are well studied and they are solved for any spacetime dimensions \cite{Iliesiu:2018fao, Alday:2020eua} thanks to the possibility to employ the method of images, which allows writing the thermal propagator as an infinite sum over zero-temperature propagators. When restricted to zero spatial coordinates this infinite sum can be recognized as a linear combination of Hurwitz $\zeta$-functions\footnote{The Hurwitz $\zeta$-function is defined as \begin{equation}
    \zeta_{H}(s,a) = \sum_{n = 0}^\infty \frac{1}{(n+a)^s} \ .
\end{equation}}
\begin{equation}\label{eq:GFFsolution}
    \langle \phi(\tau) \phi(0) \rangle_\beta = \sum_{m = -\infty}^\infty \frac{1}{(\tau+ m \beta)^{2\Delta_\phi}} = \frac{1}{\beta^{2 \Delta_\phi}} \left[\zeta_H\left(2 \Delta_\phi, \frac{\tau}{\beta}\right)+\zeta_H\left(2 \Delta_\phi, 1-\frac{\tau}{\beta}\right)\right] \ .
\end{equation}
There are cases in which the above solution simplifies even more. If we consider the GFF theory with $\Delta_\phi = 1$,
\begin{equation}\label{D=1GFF}
    \langle \phi(\tau) \phi(0) \rangle_\beta = \frac{\pi^2}{\beta^2} \csc^2\left( \frac{\pi \tau}{\beta}\right) \ .
\end{equation}
This two-point function is particularly interesting since it also corresponds to the two-point functions of a four-dimensional fundamental free scalar $\phu_{\text{4d}}$: recalling
\begin{equation}
    \langle \phu_{\text{4d}}(\tau,r)\phu_{\text{4d}}(0,0) \rangle_\beta = \frac{\pi}{2 \beta r } \left[\coth\left(\frac{\pi}{\beta} (r+ i\tau )\right)+\coth\left(\frac{\pi}{\beta} (r- i\tau )\right)\right]\  ,
\end{equation}
then 
\begin{equation}
    \langle \phi(\tau) \phi(0) \rangle_\beta=\lim_{r\rightarrow 0} \langle \phu_{\text{4d}}(\tau,r)\phu_{\text{4d}}(0,0) \rangle_\beta \ .
\end{equation}
The two-point function \eqref{D=1GFF} also corresponds to the restriction to $r=0$ of the energy-energy two-point function $\Braket{\epsilon(\tau, r) \epsilon(0,0)}_{\beta}$ in the critical two-dimensional Ising model
\begin{equation}
    \langle \phi(\tau) \phi(0) \rangle_\beta=\lim_{r\rightarrow 0} \Braket{\epsilon(\tau, r) \epsilon(0,0)}_{\beta} \ .
\end{equation}
Another special case is given by the GFF with $\Delta_\phi = 2$. In this scenario, the solution \eqref{eq:GFFsolution} reduces to 
\begin{equation}\label{eq:GFFD=2}
    \langle \phi(\tau) \phi(0) \rangle_\beta = \frac{\pi^4}{3\beta^4} \left(2+\cos\left(\frac{2\pi \tau}{\beta}\right)\right)\csc^4 \left(\frac{\pi}{\beta} \tau\right) \ .
\end{equation}
Similarly to the previous one, this case can also be interpreted as the restriction of a higher dimensional free theory to one dimension by setting $r=0$. This is indeed the case for the two-point function of a six-dimensional fundamental free scalar $\phu_{\text{6d}}$
\begin{multline}\label{eq:6dfreetheory}
     \langle \phu_{\text{6d}}(\tau,r) \phu_{\text{6d}}(0) \rangle_\beta =  \frac{\pi}{4 \beta^2 r^3} \left[\beta \coth \left(\frac{\pi  (r-i \tau )}{\beta }\right)+\beta \coth \left(\frac{\pi  (r+i
   \tau )}{\beta }\right)+\right. \\ \left . \pi  r \ \text{csch}^2\left(\frac{\pi  (r-i \tau )}{\beta
   }\right)+\pi r\ \text{csch}^2\left(\frac{\pi  (r+i \tau )}{\beta }\right)\right] \ .
\end{multline}
Equation \eqref{eq:6dfreetheory} reduces to equation \eqref{eq:GFFD=2} in the limit $r \to 0$. 
GFF theories represent an optimal playground to test the equation \eqref{eq:BootstrapFor1D} since the OPE coefficients $a_{\Delta}$ can be easily extracted from the exact solution \eqref{eq:GFFsolution}. The numerical tests for GFF theories with $\Delta_\phi = 1$ and $\Delta_\phi = 2$ are presented in Figs. \ref{fig:D1GFF} and \ref{fig:D3GFF}.
What we immediately learn from these simple checks is that, even if the equations \eqref{eq:BootstrapFor1D} are satisfied for every odd value of $\ell$, the bigger $\ell$ is, the more operators we need to sum on the right-hand side of equation \eqref{eq:BootstrapFor1D} to converge to the correct value, or better to the \emph{truncated sum} on the right-hand side to be a good approximation.  This makes the sum rules \eqref{eq:BootstrapFor1D} extremely difficult to solve by truncation of the right-hand side,  since the more OPE coefficients $a_{\Delta}$ we want to extract, the more operators we need to take into consideration. At this stage this is an observation, nonetheless in Section \ref{sec:tauberianOPE} we will show a general result about heavy operators, and that the observations made in this Section for specific examples are general (see in particular Section \ref{sec:error}).
\newline

\begin{figure}[h!]
\begin{subfigure}[t]{.49\textwidth}
  \centering
  \includegraphics[width=\textwidth]{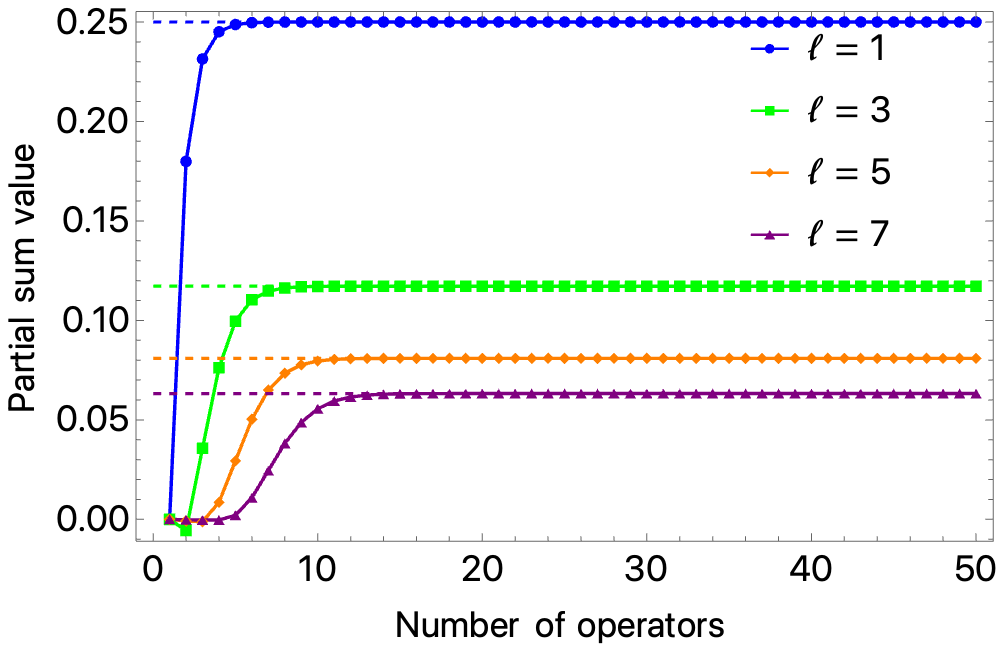}
  \caption{}
  \label{fig:Ising2dcheck}
\end{subfigure}%
\hfill
\begin{subfigure}[t]{.49\textwidth}
  \centering
  \includegraphics[width=\linewidth]{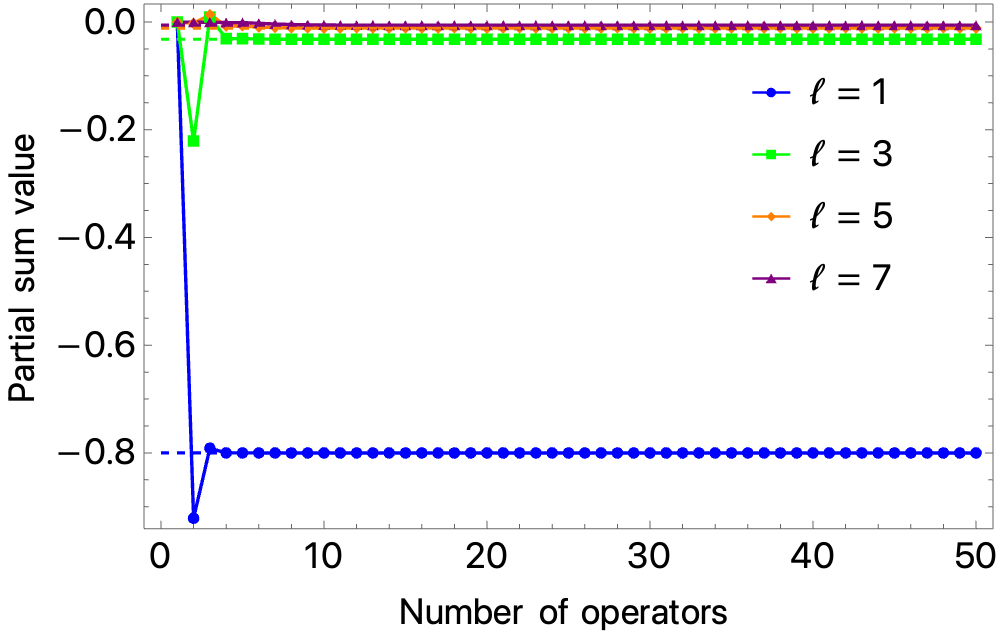}
  \caption{}
  \label{fig:YL2dCheck}
\end{subfigure}
\label{fig:2dCheck}
\caption{\emph{Comparing left (straight, dashed lines) and right (continuous lines) side of equation \eqref{eq:BootstrapFor1D} for different values of $\ell$. Observe that similarly to Figs. \ref{fig:D1GFF} and \ref{fig:D3GFF}, the bigger $\ell$ is, the more operators we need to sum for convergence. The \textbf{left panel (a)} corresponds to a numerical test for the $\langle \sigma(\tau)\sigma(0)\rangle_\beta$ correlator in the \underline{two-dimensional critical Ising model}. The \textbf{right panel (b)} corresponds to a numerical test for the $\langle \phi(\tau) \phi(0)\rangle_\beta$ correlator in the \underline{Lee-Yang model}, i.e. the non-unitary minimal model $\mathcal M_{2,5}$. In both cases, it was convenient to multiply both sides of equation \eqref{eq:BootstrapFor1D} by a factor $1/\ell !$.}}
\end{figure}

\paragraph{Second test: two-point functions of Virasoro primaries}
Another test we can easily perform concerns the two-dimensional correlators between Virasoro primary fields. The two-point functions can be computed by considering the conformal map between the plane and the cylinder \cite{DiFrancesco:1997nk,Mussardo:2020rxh,Datta:2019jeo} 
\begin{equation}
    \langle \phi (\tau,\sigma) \phi(0,0)\rangle_\beta = \left(\frac{\pi}{\beta}\right)^{2 h_\phi+2\overline h_\phi} \operatorname{csch}^{2h_\phi}\left[\frac{\pi}{\beta}(\sigma+i \tau)\right]\operatorname{csch}^{2\overline h_\phi}\left[\frac{\pi}{\beta}(\sigma-i \tau)\right]\ ,
\end{equation}
where $(h_\phi,\overline h_\phi)$ are the conformal weights of the Virasoro primary $\phi$ of conformal dimension $\Delta = h_\phi+\overline h_\phi$. When reduced to zero spatial coordinates the correlator above is given by 
\begin{equation}
      \langle \phi (\tau,0) \phi (0,0)\rangle_\beta  = \left(\frac{\pi}{\beta}\right)^{2\Delta} \csc^{2\Delta}\left(\frac{\pi \tau}{\beta}\right) \ .
\end{equation}
We can therefore test the equation \eqref{eq:BootstrapFor1D} for any dimension $\Delta$. The numerical tests are presented in Figs. \ref{fig:Ising2dcheck} and \ref{fig:YL2dCheck}: similarly to \ref{fig:D1GFF} and \ref{fig:D3GFF}, they compare the left and right-hand sides of the equation \eqref{eq:BootstrapFor1D}, for different values of $\ell$ and different levels of truncation of the sum. The considered examples are $\langle \sigma(\tau) \sigma(0)\rangle_\beta$, where the field $\sigma$ is the Virasoro primary of conformal weights $(1/16,1/16)$ in the two-dimensional critical Ising model, and $\langle \phi(\tau)\phi(0)\rangle_\beta$, where $\phi$ is the only Virasoro primary field in the Lee-Yang minimal model. The Lee-Yang model is the simplest possible conformal theory since it is the minimal model with only one Virasoro primary, of conformal weights $(-1/5,-1/5)$. Despite its simplicity, the Lee-Yang model is not unitary: this allows us to stress that the only assumption we used to derive equation \eqref{eq:BootstrapFor1D} was the KMS condition, and therefore the sum rules \eqref{eq:BootstrapFor1D} hold also for non-unitary theories, such as Lee-Yang.
\begin{figure*}[t]
\centering
   \includegraphics[width=0.55\textwidth]{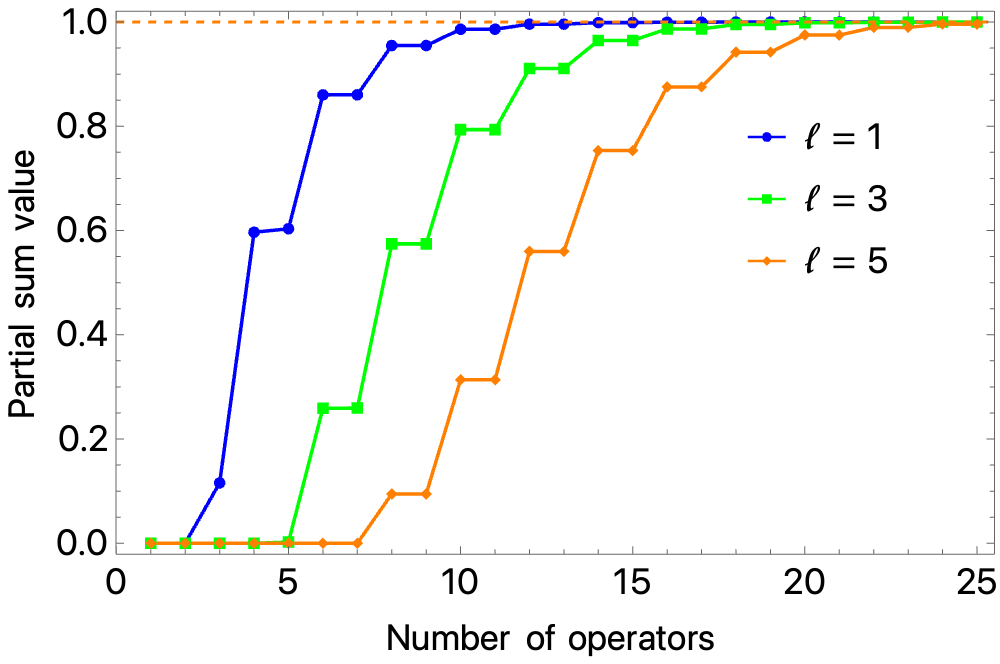}
  \caption{\emph{Comparing left (straight, dashed lines) and right (continuous lines) side of equation \eqref{eq:BootstrapFor1D} for different values of $\ell$ for the two point function of fundamental scalars in the \underline{3d $\mathrm O(N)$ model at large $N$}. The explicit solution of the two-point function at zero spatial coordinate is given in Section \ref{eq:O(N)largeN}: the operators contributing in the sum rules are powers of the Hubbard-Stratonovich field $\sigma^m$ and the double twists of the fundamental field $[\phi_i\phi_i]_{n,l}$. It was convenient to multiply both sides of equation \eqref{eq:BootstrapFor1D} by a factor $1/\ell !$. }}\label{figtauberiancheck1}
\end{figure*}
\subsection{Light operators' one-point functions} \label{ssec: larggap}
As observed in the numerical tests performed in Section \ref{ssec: zero}, the sum rules \eqref{eq:BootstrapFor1D} are difficult to solve by OPE truncation. Indeed, by looking at the numerical tests \ref{fig:D1GFF}, \ref{fig:D3GFF}, \ref{fig:Ising2dcheck}, \ref{fig:YL2dCheck}, usually the number of equations needed to produce a good approximation, labeled by $\ell$,  is lower than the number of operators in the truncated OPE. If this happens, the number of equations is lower than the number of coefficients $a_{\Delta}$, and therefore the system of equations is not determined.  In general, it is difficult to know \emph{a priori} how many equations are needed for a correct approximation, i.e. given an OPE truncation it is difficult to find a cutoff $\widehat \ell$ such that all the equations produced by $1 \leq \ell < \widehat \ell$ can be considered correct in the approximation  (see Section \ref{sec:error} for a quantitative estimation). This makes the bootstrap problem defined by the set of equations \eqref{eq:BootstrapFor1D} not solvable.
\newline Despite the comment above, there is a class of theories for which the set of equations \eqref{eq:BootstrapFor1D} can be used. This happens when the OPE decomposes in a light operators' sector, made by a \emph{finite} number of operators, and in a heavy operators' sector. The OPE can be truncated so that only the light operators contribute and the heavy ones can be neglected. In this scenario, we choose $\ell$ equations so that we can approximately solve the bootstrap problem. In the following, we solve the problem for one and two non-trivial (i.e. different from the identity) light operators, estimating their one-point functions. 
\newline
When the sum rules in equation \eqref{eq:BootstrapFor1D} are truncated, an approximation \emph{error} is inevitably introduced.  We will employ the Tauberian theorem (see Section \ref{sec:tauberianOPE}) to estimate it quantitatively.
\paragraph{One light operator}
The simplest case is when in the OPE between two fields $\phi$ contains one light operator in addition to the identity operator $\mathds{1}$, namely $\mathcal O_{1}$
\begin{equation}
    \phi(\tau) \times \phi(0) \sim \mathds{1} +  \mathcal O_{1}+ \text{heavy operators} \ .
\end{equation}
If we assume that all the heavy operators' contributions can be neglected, then the only equation we need is ($\ell=1$)
\begin{equation}
    2 \Delta_\phi \simeq \frac{a_{\Delta_1}}{2^{\Delta_{1}}} \left(\Delta_{1}-2\Delta_\phi\right) \ ,
\end{equation} 
which can be solved for the one-point function thermal OPE coefficient $b_{\mathcal O_L}$
\begin{equation}
    b_{\mathcal O_{1}}\simeq 2^{1+\Delta_{1}}\frac{c_{\mathcal O_{1}}}{f_{\phi \phi \mathcal O_{1}}}\frac{\Delta_\phi}{\Delta_{1}-2\Delta_\phi}  \ .
\end{equation}

\paragraph{Two light operators}
The next simple case is the one in which only the identity plus two light operators $\mathcal{O}_{1}$ and $\mathcal{O}_{2}$ are considered, while all the others are heavy enough to be negligible. We are therefore assuming an OPE of the form 
\begin{equation}
    \phi(\tau) \times \phi(0) \sim \mathds{1} +  \mathcal O_{1}+\mathcal{O}_2+ \text{heavy operators} \ ,
\end{equation}
from which we can use two equations of the form \eqref{eq:BootstrapFor1D} corresponding to $\ell = 1$ and $\ell =3$. In particular, the two equations to solve are 
\begin{equation}
      2 \Delta_\phi \simeq \frac{a_{\Delta_1}}{2^{\Delta_{1}}} \left(\Delta_{1}-2\Delta_\phi\right)+ \frac{a_{\Delta_2}}{2^{\Delta_{2}}} \left(\Delta_{2}-2\Delta_\phi\right) \ , 
\end{equation}
and \begin{equation}
      4 \Delta_\phi (1+\Delta_\phi)(1+2\Delta_\phi) \simeq \frac{a_{\Delta_1}}{2^{\Delta_{1}}} \frac{\Gamma(\Delta_{1}-2\Delta_\phi+1)}{\Gamma(\Delta_{1}-2\Delta_\phi-2)}+\frac{a_{\Delta_2}}{2^{\Delta_{2}}} \frac{\Gamma(\Delta_{2}-2\Delta_\phi+1)}{\Gamma(\Delta_{2}-2\Delta_\phi-2)} \ .
\end{equation}
Solving the two equations for the thermal OPE coefficients $b_{{\mathcal{O}_{1}}}$ and $b_{{\mathcal{O}_{2}}}$, we get 
\begin{equation}
    b_{\mathcal O_{1}} \simeq 2^{1+\Delta_1} \frac{c_{\mathcal O_{1}}}{f_{\phi \phi \mathcal O_{1}}}\frac{\Delta_\phi}{\Delta_1-2\Delta_\phi} \left[\frac{ \Delta_2 (\Delta_2-4\Delta_\phi-3)}{(\Delta_2-\Delta_1)(\Delta_1+\Delta_2-4\Delta_\phi-3)}\right] \ ,
\end{equation}
\begin{equation}
    b_{\mathcal O_{2}} \simeq 2^{1+\Delta_2} \frac{c_{\mathcal O_{2}}}{f_{\phi \phi \mathcal O_{2}}}\frac{\Delta_\phi}{\Delta_2-2\Delta_\phi} \left[\frac{ \Delta_1 (\Delta_1-4\Delta_\phi-3)}{(\Delta_1-\Delta_2)(\Delta_1+\Delta_2-4\Delta_\phi-3)}\right] \ .
\end{equation}
\newline Including more light operators is straightforward and the linear system can be solved. The case where two light operators have the same conformal dimension is different. In these case, equation \eqref{eq:BootstrapFor1D} is not sufficient to solve for one-point functions and we need to consider the full set of equations in \eqref{eq:sumrule}. Furthermore, generically the error of the truncation increases with the number of light operators since more and more equations are needed in order to solve the theory.
\paragraph{Constraining zero-temperature CFT data from KMS sum rules}
In the previous Section, we employed the equations \eqref{eq:BootstrapFor1D} to extract thermal one-point functions $b_{\mathcal{O}}$. Nevertheless, the same equations also contain information about the zero-temperature CFT data, such as conformal dimensions $\Delta_{\mathcal{O}}$ and the structure constants $f_{\phi \phi \mathcal{O}}$; if we consider \eqref{eq: sumrules}, the sum rules also contain information on the spins $J_{\mathcal{O}}$. If we assume that the contribution from heavy operators can be neglected in the right-hand side of the equation \eqref{eq:BootstrapFor1D}, it is possible to constrain the conformal spectrum for the light operators sector. To show this, let us consider a theory with a light sector composed of $N$  light operators $\mathcal{O}_{1}, \dots \mathcal{O}_N$: the sum rule \eqref{eq:BootstrapFor1D} can be approximated (see Section \ref{sec:error} for a quantitative estimation of the approximation) by
\begin{equation}\label{eq:largegapped eq}
    \sum_{i = 1}^N a_{\Delta_{i}} F(\Delta_{i},2p-1) \simeq 0 \ , \qquad F(\Delta_{i},2p-1) = \frac{\Gamma(\Delta_{i}-2\Delta_\phi+1)}{2^{\Delta_{i}}\Gamma(\Delta_{i}-2\Delta_\phi-2p+2)} \ ,
\end{equation}
where $p = 1,\ldots, N$; the identity contribution has been moved to the left side of this equation, i.e. $\mathcal O_1 = \mathds{1}$. The equation \eqref{eq:largegapped eq} is a set of $N$ linear homogeneous equations in $N$ variables. We require the system to have a physical solution, i.e. a set of non-trivial thermal OPE coefficients $ a_{\Delta_{1}}, \dots,  a_{\Delta_{N}} $: if we consider $F(\Delta_{i},2p-1)$ as a square matrix with columns labeled by $i=1, \dots, N$ and rows labeled by $p=1, \dots, N$, then we must require its determinant to be zero
\begin{equation} \label{eq: det}
    \det \big[F(\Delta_{i},2p-1)\big] = 0  \ .
\end{equation}
This is a non-trivial equation for the conformal dimensions $\Delta_{1}, \dots, \Delta_{N}$ of the light operators. As a concrete example, we consider the case of two light operators $\mathcal{O}_{1}=\mathds{1}$ and $\mathcal{O}_2$: the equation \eqref{eq: det} implies that the light conformal dimensions are $\Delta_{\mathds{1}}=0$, and $\Delta_{2} \in \{ 2\Delta_\phi, 4\Delta_\phi+3\}$.
\newline To our best knowledge, this approach was never explored before in the context of thermal CFTs, but similar comments are present in the literature for zero temperature CFTs when considering the crossing equation instead of equation \eqref{eq:BootstrapFor1D}. Interestingly, this strengthens the analogy between the KMS condition and crossing symmetry (for finite and zero temperature respectively), proposed by \cite{El-Showk:2011yvt} and stressed in \cite{Iliesiu:2018fao,Iliesiu:2018zlz}. 

The idea of using a truncated OPE goes under the name of \textit{Gliozzi method}, introduced in \cite{Gliozzi:2013ysa}, and it is used in different cases, including non-unitary theories \cite{Gliozzi:2014jsa, Gliozzi:2015qsa, Gliozzi:2016cmg, Nakayama:2016cim, Nakayama:2021zcr}. Both the truncated equations presented here and the zero-temperature truncated crossing equations produce relations between conformal dimensions of light operators in the OPE. Nonetheless, there is not necessarily a correspondence between the theories that can be studied at zero-temperature by using the Gliozzi method (i.e. theories with \textit{truncatable bootstrap equations}) and theories that can be studied at finite temperature by truncating the equations \eqref{eq:BootstrapFor1D}: an example is brought by the two-dimensional critical Ising model. We currently do not know if there are theories in which the truncated version of \eqref{eq:BootstrapFor1D} returns conformal dimensions that cannot be computed from the truncated crossing equation at zero-temperature. Finally, it would be interesting to extend this notion of \textit{truncability} also for the most general case \eqref{eq:sumrule}.
\section{A Tauberian theorem for thermal CFTs}\label{sec:tauberianOPE}
In this Section, we delve into how the divergences of the two-point correlation function, together with the KMS condition, can be used to gain insights into the thermal OPE coefficients when the spatial coordinates are fixed to zero. This result can be derived by invoking the \emph{Tauberian theory} (see \cite{tauberian} for an extensive review). Tauberian theory is a mathematical theory that found application in physics first in \cite{Vladimirov:1978xx}, and more recently in \cite{Pappadopulo:2012jk,Das:2017vej,Mukhametzhanov:2019pzy,Pal:2019zzr,Pal:2019yhz,Ganguly:2019ksp,Das:2020uax,Mukhametzhanov:2020swe,Pal:2020wwd,Kusuki:2023bsp,Pal:2023cgk}. Notably, its application to one-dimensional CFTs is rigorously detailed in \cite{Qiao:2017xif}. Motivated by these works, we aim to derive the thermal OPE density for heavy operators by combining a Tauberian theorem with the KMS condition and the parity transformation $\tau \to -\tau$ on the thermal circle. 

\noindent We first present a heuristic derivation of such theorem to provide an intuitive sense of its meaning and outcome (see Section \ref{ssec: deriv}). Following this, we will outline the steps required to establish it rigorously. The crucial hypothesis of the theorem, i.e. the boundedness of the OPE coefficients, will be proven by showing that the OPE coefficients corresponding to heavy operators all share the same sign (from a certain cutoff on). Once proven, the Tauberian theorem can serve multiple purposes. First, it allows us to compute the leading order and the leading term of the thermal OPE coefficients $a_{\Delta}$ for heavy operators (see Section \ref{ssec: heavy}). Moreover, it can be used to approximate and bound the thermal two-point functions of identical, scalar operators (see Section \ref{ssec: twopo}). We will also comment on the possibility of using a similar technology to study the thermal two-point functions of non-identical operators. Finally, as anticipated in Section \ref{sec:KMSTCFT}, the Tauberian theorem can be used to complete the discussion about the truncability of the sum rules \eqref{eq:sumrule} by explicitly estimating the approximation error (see Section \ref{sec:error}). Many examples are shown for free theories, two-dimensional models, and for the three-dimensional $\mathrm O(N)$ model at large $N$. 
\subsection{Derivation of the theorem} \label{ssec: deriv}
Here we present the derivation of the thermal Tauberian theorem. First, we will present a heuristic derivation that will clear the physics and then we will justify it.
\subsubsection{Heuristic derivation}  \label{ssec: heur}
\begin{figure}[h!]
\centering
\begin{tikzpicture}
[x=0.6pt,y=0.6pt,yscale=-1,xscale=1]
\draw[->,thick] (-200,0)--(200,0);
\draw (155,-70)--(155,-55);
\draw (155,-55)--(175,-55);

\filldraw[red] (0,0) circle (2pt) node[anchor=north]{{\color{black}{$0$}}};
\filldraw[red] (50,0) circle (2pt) node[anchor=north]{{\color{black}{$\beta$}}};
\filldraw[red] (100,0) circle (2pt) node[anchor=north]{{\color{black}{$2 \beta$}}};
\filldraw[red] (150,0) circle (2pt) node[anchor=north]{{\color{black}{$3 \beta$}}};
\filldraw[red] (-50,0) circle (2pt) node[anchor=north]{{\color{black}{$-\beta$}}};
\filldraw[red] (-100,0) circle (2pt) node[anchor=north]{{\color{black}{$-2\beta $}}};
\filldraw[red] (-150,0) circle (2pt) node[anchor=north]{{\color{black}{$-3\beta$}}};

\draw (160,- 70) node [anchor=north west][inner sep=0.75pt]  [font=\large]  {$\tau$};
\end{tikzpicture}
\caption{\emph{Location of the poles of the function $\langle \phi(\tau)\phi(0)\rangle_\beta$ on the $\tau$ real axis. Although they do not appear in this picture, the presence of possible branch cuts is allowed.}}
\label{PolesInTwoPointFunctions}
\end{figure}
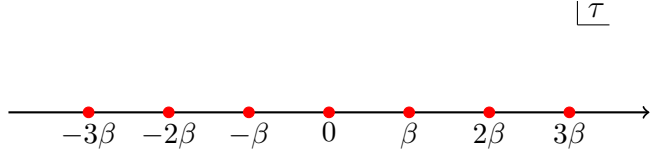

Let us consider a scalar, local operator $\phi(\tau, \Vec{x})$ and its thermal two-point function 
\begin{equation}
    \langle \phi(\tau,r) \phi(0,0)\rangle_\beta \ ,
\end{equation}
and consider the same two-point function in the limit of zero spatial coordinates
\begin{equation}
       \langle \phi(\tau) \phi(0)\rangle_\beta \equiv  \langle \phi(\tau,0) \phi(0,0)\rangle_\beta \ .
\end{equation}
In the same limit, the thermal OPE expansion of the two-point function
\begin{equation}\label{eq:TwoPointFunctionInt}
    \langle \phi(\tau) \phi(0)\rangle_\beta = \tau^{-2 \Delta_\phi}\sum_{\mathcal O \in \phi \times \phi} \frac{a_{\mathcal O}}{\beta^{\Delta_{\mathcal O}}} \tau^{\Delta_{\mathcal O}} = \tau^{-2\Delta_\phi} \int_0^\infty \de \Delta \ \rho(\Delta) x^{\Delta} \ ,
\end{equation}
where we introduced the \emph{spectral density}  $\rho(\Delta)$
\begin{equation}\label{eq:densitydef}
    \rho(\Delta) = \sum_{\mathcal O \in \phi \times \phi} a_{\mathcal O} \ \delta(\Delta-\Delta_{\mathcal O}) \ ,
\end{equation} 
and the dimensionless ratio $x = \tau/\beta$. The thermal OPE does not converge everywhere, but only in the open interval $0< x < 1$ \cite{Iliesiu:2018fao}: this is due to the presence of an infinite number of poles on the real $\tau$ axis (see Fig.\ref{PolesInTwoPointFunctions}). 
We can easily convince ourselves of this by observing that the KMS condition implies that 
\begin{equation}
    \langle \phi(\tau) \phi(0)\rangle_\beta \overset{\tau \to k \beta}{\sim} (k\beta-\tau)^{-2\Delta_\phi} \ ,
\end{equation}
for any $k \in \mathbb Z$. Hence, recalling the equation \eqref{eq:TwoPointFunctionInt} we can write  \begin{equation}
    \int_0^\infty \de \Delta \  \rho(\Delta) x^{\Delta}\  \overset{x \to 1}{\sim}\  (1-x)^{-2\Delta_\phi} \ .
\end{equation}
Now, we make the \emph{ansatz} that the behavior of $\rho(\Delta)$  in the limit  $\Delta \to \infty$ is a power-law behavior
\begin{equation}\label{eq:assumption}
    \rho(\Delta) \overset{\Delta \rightarrow \infty}{\sim} A \Delta^{-\alpha+1} \ .
\end{equation} 
This ansatz is justified in Section \ref{ssec: precise}. Plugging it in the two-point function \eqref{eq:TwoPointFunctionInt} we find
 \begin{equation} \label{eq: taubeta}
    \langle \phi(\tau) \phi(0) \rangle_\beta \overset{\tau \to \beta}{\sim}  A \, \Gamma(2-\alpha ) \tau^{-2 \Delta_\phi } \left(1-\frac{\tau}{\beta}\right)^{\alpha-2} \ ,
\end{equation}
from which we can read \begin{equation}
   \alpha = -2\Delta_\phi+2 \ , \qquad A = \frac{1}{\Gamma(2\Delta_\phi)}  \ .
\end{equation}
Therefore, the behavior of $\rho(\Delta)$ in the limit $\Delta \rightarrow \infty$ becomes  
\begin{equation}\label{eq:Tauberian}
   \rho(\Delta) \overset{\Delta \rightarrow \infty}{\sim}   \frac{1}{\Gamma(2 \Delta_\phi)} \Delta^{2 \Delta_\phi-1}\ .
\end{equation}
The physical interpretation of this result is in terms of the density, in the $\phi \times \phi$ OPE, of heavy operators (captured by the limit $\Delta \rightarrow \infty$). Let us observe that the formula \eqref{eq:Tauberian} should not be regarded as physical: indeed, the conformal spectrum could be discrete, and therefore the density could, in general, be a sum of $\delta$-functions. Nevertheless, the formula \eqref{eq:Tauberian} acquires a correct physical meaning when considered ``on average''
\begin{equation}\label{eq:TauberianMath}
    \int_0^{\Delta} \rho(\widetilde \Delta) \ \de \widetilde \Delta \ \overset{\Delta \to \infty}{\sim} \frac{\Delta^{2 \Delta_\phi}}{\Gamma(2 \Delta_\phi+1)} \ .
\end{equation}
\subsubsection{The precise derivation} \label{ssec: precise}
The above derivation is correct once the ansatz expressed in \eqref{eq:assumption} is proved, i.e. assuming that the density of states grows as a power-law behavior of the conformal dimension $\Delta$ when $\Delta \rightarrow \infty$. As already commented above, this statement must be translated mathematically in its ``integrated version'', as shown in the equation \eqref{eq:TauberianMath}. The proof is non trivial and we have to invoke the \textit{Tauberian theory}. First of all, let us highlight that the assumption we made is not general: a standard counterexample is the series \cite{tauberian2,Qiao:2017xif}
\begin{equation}\label{eq:contr}
    \frac{1}{(x+1)^2(1-x)}= 1-2x+2 x^2-3 x^3+3x^4+\ldots \ .
\end{equation}
Indeed, this series asymptotically behaves as expected, since for $x \to 1$ it grows as $(x-1)^{-1}$, but the partial sums of the series coefficients are given by $1,0,2,0,3,0,\ldots$, hence they end up oscillating between 0 and $n$ for $n \rightarrow \infty$ preventing a power-law description of the partial sums. A more general problematic situation was proposed in \cite{Qiao:2017xif}, where the authors observed that a possible behavior of the partial sums, without any other assumption, can be an oscillation between two power-law behaviors. This problem was rigorously faced by mathematicians in the last century and the results go under the name of \textit{Tauberian theorems}. In \cite{Pappadopulo:2012jk} and \cite{Qiao:2017xif} the key assumption is the hypothesis that the weighted density $\rho(\Delta)$ is non-negative. However, this is not correct in our case: recalling the definition of our density \eqref{eq:densitydef}, the coefficients $a_{\mathcal O}$ are allowed to be negative. The lack of positivity of the OPE coefficients is known to be one of the main obstacles in the thermal conformal bootstrap program. 

\noindent To approach this problem and prove rigorously a thermal Tauberian theorem, we want first to relax the positivity hypothesis. Indeed, a version of the Tauberian theorems can be proved using three sufficient conditions, instead of positivity: 
\begin{itemize}
    \item[$\star$] reality of the OPE coefficients $a_{\mathcal O}$;
    \item[$\star$] boundedness from below of the OPE coefficients $a_{\mathcal O}$;
    \item[$\star$] $\Delta_\phi > \frac{1}{2}$.
\end{itemize}
 For a precise derivation of the Tauberian theorem under these hypotheses, see the  Appendix \ref{appendix:OPEdensityProof}, or the Theorems 95 and 97 in \cite{Hardy}. Our approach will now consist of showing that the three sufficient hypotheses apply to our case. The last assumption ensures that $\alpha > 1$, where $\alpha$ controls the divergence $(x-1)^{-\alpha}$ in the limit $x \to 1$. This is important to the theorem to be true as clarified in Appendix \ref{appendix:boundedness}. This assumption could be physically motivated by the unitarity bound for scalars
 \begin{equation}
    \Delta_\phi \ge \frac{d-2}{2} \ ,
\end{equation}
which means that the only unitary, physically relevant theory that does not satisfy this bound (in $d >2$) is the free scalar theory in three dimensions (which saturates the bound). This theory can be solved analytically as explained in the previous Section so we do not worry about this case. Regarding two-dimensional theories, the conformal dimensions of scalars do not have to be greater than $1/2$. Indeed,  this case can be solved analytically as well by using the conformal map between the plane and the cylinder. Therefore, the statements we are going to make apply to non-trivial and generically not-solvable unitary physical models.
\newline  The reality of the OPE coefficients $a_{\mathcal{O}}$ is physically motivated by unitarity. Hence, we are only required to prove the boundedness from below of these coefficients. We employ the following technique: we will prove that there exists a cutoff conformal dimension $\widehat{\Delta}$ such that, for $\Delta > \widehat{\Delta}$, the thermal OPE coefficients all share the same sign (say positive\footnote{In the case in which the sign is negative (e.g. the Lee-Yang model), the same procedure can be adopted by flipping the sign of the density.}). Since each thermal OPE coefficient is finite (to ensure the convergence of the OPE expansion), the boundedness from below will come as a corollary of the same-sign statement, with the bound from below given by the lowest OPE coefficient among the operators with conformal dimension $0 < \Delta < \widehat{\Delta}$. 
\begin{figure}
    \centering
\begin{tikzpicture}
\fill[color={AliceBlue}] (0,0) rectangle (4.38,5.3);
\fill[color={MistyRose}]  (4.38, 0) rectangle (7.2,5.3);
\node[align=left] at (0.6,4.7) {\textcolor{blue}{\emph{Light}} \\
\textcolor{blue}{\emph{sector}}};
\node[align=left] at (4.98,4.7) {\textcolor{red}{\emph{Heavy}} \\
\textcolor{red}{\emph{sector}}};
\begin{axis}[thick, 
    width=9cm,
    height=7cm,
    axis lines=middle,
    ylabel=$a_{\Delta}$,
    xlabel=$\Delta$,
    xmin=0,
    ymin=-4,
    xmax=11,
    ymax=10,
    xtick={100},
    ytick={100},
    grid=none,
    ticklabel style={font=\normalsize},
    xlabel style={at={(current axis.right of origin)},anchor=west, },
    ylabel style={at={(current axis.above origin)},anchor=south},
]
\addplot+ [nodes near coords, color=blue, only marks, point meta=explicit symbolic, mark=*,mark options={fill=blue} ,  coordinate style/.from=\thisrow{style}]
table [meta=label] {
x y label style
1 1 $\mathcal{O}_1$ {}
2.5 -2 $\mathcal{O}_2$ below
4 4 $\mathcal{O}_3$ {}
4.5 -2.5 $\mathcal{O}_4$ below
6 2.5 $\mathcal{O}_5$ {}
};
\addplot+ [nodes near coords, color=red, only marks, point meta=explicit symbolic, mark=*,mark options={fill=red} ,coordinate style/.from=\thisrow{style}]
table [meta=label] {
x y label style
7.5 3.5 $\mathcal{O}_6$ {}
9 5 $\mathcal{O}_7$ {}
10 6.5 $\mathcal{O}_8$ {}
};
\addplot+ [nodes near coords, color=Green, only marks, point meta=explicit symbolic, mark=square*, mark options={fill=Green},coordinate style/.from=\thisrow{style}]
table [meta=label] {
x y label style
6.5 0  {} {}
};
\addplot[black, dashed, no markers]coordinates {(1,1) (1,0)};
\addplot[black, dashed, no markers]coordinates {(2.5,-2) (2.5,0)};
\addplot[black, dashed, no markers]coordinates {(4,4) (4,0)};
\addplot[black, dashed, no markers]coordinates {(4.5,-2.5) (4.5,0)};
\addplot[black, dashed, no markers]coordinates {(6,2.5) (6,0)};
\addplot[black, dashed, no markers]coordinates {(7.5,3.5) (7.5,0)};
\addplot[black, dashed, no markers]coordinates {(9,5) (9,0)};
\addplot[black, dashed, no markers]coordinates {(10,6.5) (10,0)};
\addplot [Green, no markers, dashed, very thick] coordinates {(6.5,-12) (6.5,20)};
\end{axis}
\node[align=left] at (4.6,1.3) {\small \textcolor{Green}{$\widehat{\Delta}$}};
\node[align=left] at (0.2,1.3) {\small{$0$}};
\end{tikzpicture}
 \caption{\emph{Schematic depiction of the expected behavior of the OPE coefficients $a_{\Delta}$. The cutoff conformal dimension $\widehat{\Delta}$ is represented by a green square and dashed line, and it divides the set of all the operators in the OPE $\phi \times \phi$ into two sectors. In the ``light sector'', represented in blue, we have a finite number of operators with $\Delta < \widehat{\Delta}$, and the coefficients $a_{\Delta}$ are allowed to be both positive and negative. 
 In the ``heavy sector'', represented in red, we have an infinite number of operators with  $\Delta > \widehat{\Delta}$; in this work, it is argued that the sign of the coefficients $a_{\Delta}$ is always the same (positive in the picture). An OPE spectrum with a behavior similar to this pictorial representation ensures that the Tauberian theorem holds. As a concrete example, a quantitative plot of the OPE spectrum was produced for the $\mathrm O(N)$ model at large $N$ (see Fig. \ref{fig:ONan}).}}
    \label{fig:ExpectedBev}
\end{figure}
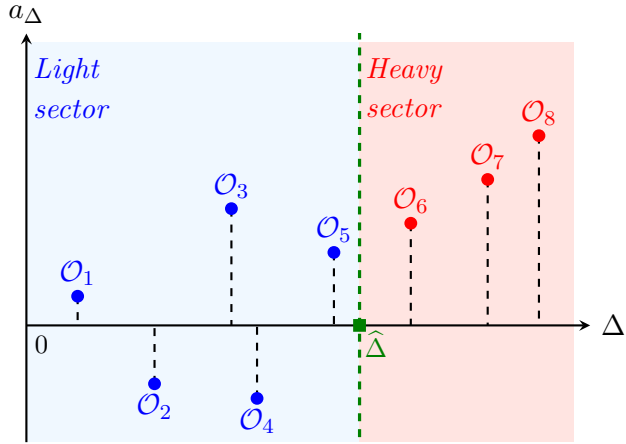
 A pictorial representation of the behavior of the coefficients $a_{\Delta}$ as a function of $\Delta$ is shown in Fig. \eqref{fig:ExpectedBev}; a concrete example of such behavior is instead shown in Fig. \ref{fig:ONan}, in the case of the $\mathrm O(N)$ model. As already anticipated, we do not expect such representation to be universal (as shown by the counterexample \eqref{eq:contr}), but rather to be a consequence of the KMS condition. The physical intuition is the following: 
 since each term in the OPE expansion does not present any singularity when $\tau \to \beta$, the power-law singularity predicted by KMS has to be a cumulative effect of the tail of operators at $\Delta \gg 1$. The claim is that, due to this channel duality, the stability of the signs of the coefficients is ensured. The proof of the sign stability for large $\Delta$ is given in the Appendix \ref{appendix:boundedness}. In summary, we can make use of the Euclidean inversion formula combined with KMS to compute a coefficient $a_{\Delta}$ in terms of all the other OPE coefficients. We then compare $a_{\Delta}$ and $a_{\Delta+\delta \Delta}$, where $\delta \Delta$ is the minimal distance between two operators. The result for $\Delta \gg 1$  is \begin{equation}\label{eq:samesign}
    a_{\Delta+\delta \Delta}  = a_{\Delta} \left[1+ \mathcal{O}\left(\frac{1}{\Delta}\right) \right] \ ,  \quad \Delta \gg 1 \ .
\end{equation}
Then if $a_{\Delta}$ is not going to zero, there will be a cutoff $\widehat{\Delta}$ such that the correction $\mathcal{O}\left(1/ \widehat{\Delta} \right)$ becomes negligible compared to $a_{\widehat{\Delta}}$. As a consequence, all the heavy contributions $a_{\Delta}$ ($\Delta > \widehat{\Delta}$) share the same sign.
This conclusion can be explicitly checked and it is correct in all the known examples, up to our knowledge. In particular, apart from the free scalar theories and the two-dimensional theories discussed in this work, holographic computations reproduce the correct behavior. The latter computations were done by considering the geodesic approximation of a two-point function of heavy operators in \cite{Rodriguez-Gomez:2021pfh}. Furthermore, as a check, we also considered the three-dimensional $\mathrm O(N)$ model at large $N$: the two-point function is known \cite{Iliesiu:2018fao}, and the results and details are provided in Section \ref{eq:O(N)largeN}. 
\newline Let us stress that this does not imply the positivity of the OPE coefficients $a_{\Delta}$, but rather that all the OPE coefficients of the operators in the heavy sector share the same sign. They can be all negative or all positive, and this does not represent an obstacle to the derivation of the Tauberian theorem (just by multiplying by a sign factor it is possible to have a density bounded from below). In our exploration, we also checked non-unitary models such as the Lee-Yang model in two-spacetime dimensions. This is one of the cases in which the coefficients corresponding to heavy operators are all negative. This was expected since the left side of equation \eqref{eq:BootstrapFor1D} is negative for any $\ell \in 2 \mathbb N +1$ (recall that the conformal dimension of the Virasoro primary field is $\Delta = - 2/5$). This could suggest that the positivity of the heavy dimensional coefficients depends on the unitarity of the theory. It would be interesting to explore this more in the future.
\subsection{Applications of the theorem}
We present here some applications of the thermal Tauberian theorem derived in Section \ref{ssec: deriv}. In particular, in Section \ref{ssec: heavy} we compute the leading term of OPE coefficients associated to heavy operators. In Sections \ref{ssec: twopo} and \ref{sec:Bounding2pt} we use the thermal OPE density to approximate and bound the thermal two-point functions. As promised in Section \ref{ssec: larggap}, the Tauberian theorems can be used to evaluate the range of validity of the large gap approximation discussed above: we explain this point in Section \ref{sec:error}. Finally, we present a non-trivial example, namely the three-dimensional $\mathrm O(N)$ model at large $N$, in Section \ref{eq:O(N)largeN}.
\subsubsection{Thermal OPE coefficients for heavy operators}\label{ssec: heavy}  The most interesting application of the Tauberian theorem is to approximate the OPE coefficients $a_{\Delta}$ when $\Delta \to \infty$, namely for heavy operators. Recalling the exact definition of the thermal OPE density \eqref{eq:densitydef}, we can easily see that it is possible to isolate the OPE coefficient of a heavy operator of dimension $\Delta_{H}$  by integrating over the interval $\left[\Delta_{H}-\delta \Delta_H, \Delta_{H} \right]$, where $\delta \Delta$ parameterize the distance between $\Delta_{H}$ and the nearest neighborhood operator of smaller dimension appearing in the OPE
\begin{equation}
    a_{\Delta_{H}}=\int_{\Delta_{H}-\delta \Delta}^{\Delta_{H}} \rho(\Delta) \de \Delta \ , \quad \rho(\Delta)=\sum_{\mathcal O \in \phi \times \phi} a_{\mathcal O} \ \delta(\Delta-\Delta_{\mathcal O}) \ .
\end{equation}
If we employ the Tauberian theorem, in the heavy operators' regime we can adopt the Tauberian theorem \eqref{eq:TauberianMath}  \footnote{Make this statement mathematically rigorous is highly non-trivial: passing from a Tauberian theorem of the type \eqref{eq:TauberianMath} to a Tauberian theorem for small intervals is a notoriously difficult problem. For the sake of completeness, we clarify these points in Appendix \ref{app:TaubOPE}. }
\begin{equation}
    a_{\Delta_{H}} = \left (\int_{0}^{\Delta_{H}}\de \Delta -\int_{0}^{\Delta_{H}-\delta \Delta}\de \Delta  \right)\rho(\Delta) \overset{\Delta_{H} \to \infty}{\sim}  \frac{\Delta_{H}^{2\Delta_\phi-1}}{\Gamma(2\Delta_\phi)}\delta \Delta \ ,
\end{equation}
in the limit $\delta \Delta \ll \Delta_{H}$. Therefore 
\begin{equation}\label{eq:largeDcoeff}
    \boxed{a_{\Delta_{H}} \overset{\Delta_{H} \to \infty }{\sim}    \frac{\Delta_{H}^{2\Delta_\phi-1}}{\Gamma(2\Delta_\phi)}\delta \Delta} \ .
\end{equation}
Let us stress that in this derivation $\Delta_H$ is the conformal dimension of a heavy operator and $\delta \Delta$ is the difference between $\Delta_H$ and the conformal dimension of the nearest neighborhood operator of smaller conformal dimension.\footnote{We do not claim this result for any $\Delta_H$ and arbitrary $\delta \Delta$. To claim such a result one should prove a finite-size version of the Tauberian theorem, which is usually stronger than the Tauberian theorem presented here.} Equation \eqref{eq:largeDcoeff} is a prediction for \textit{any thermal CFT}: in the following we will check it only for theories which allow an analytical expression of the two-point function, i.e. $2d$ CFTs, generalized free theories and the $\mathrm O(N)$ model at large $N$.

\subsubsection{Approximating thermal two-point functions for $r=0$} \label{ssec: twopo}
The Tauberian density  \eqref{eq:Tauberian} provides a good approximation of the weighted density of states \eqref{eq:densitydef} in the heavy operators' regime, regardless of their spin. Taking inspiration from the equation \eqref{eq:TwoPointFunctionInt}, one can try to insert the Tauberian density in the integral
\begin{equation}
    \langle \phi(\tau) \phi(0)\rangle_\beta \overset{\tau \to \beta}{\sim}  \int_0^\infty \de \Delta \ \rho(\Delta) \frac{\tau^{\Delta-2 \Delta_{\phi}}}{\beta^{\Delta}} \ , \quad \rho(\Delta)\overset{\Delta \to \infty}{\sim} \frac{1}{\Gamma(2 \Delta_{\phi})} \Delta^{2\Delta_{\phi}-1}  \ .
\end{equation}
By combining the KMS condition and the parity transformation $\tau \to - \tau$, we can also employ such approximation in the regime $\tau\ll \beta$. The final result is 
\begin{equation}\label{eq:2ptfunctionapprox}
    \langle \phi(\tau) \phi(0) \rangle_\beta \simeq \left\{\begin{matrix}
 \displaystyle \left[(\beta-\tau)\log\left(1-\frac{\tau}{\beta} \right)\right]^{-2 \Delta_{\phi}}  & \tau/\beta \ll 1 \\ 
 \\ 
  \displaystyle \left[\tau \log\left(\frac{\tau}{\beta} \right)\right]^{-2 \Delta_{\phi}} & \tau/\beta \sim 1
\end{matrix}\right. \ .
\end{equation}
Of course, the Tauberian density can approximate the two-point function in the vicinity of its divergences, but not close to $\tau \sim \beta/2$. This is because the Tauberian density encodes information about heavy operators (when $\tau \sim \beta$) or very light operators (in particular the identity $\mathds{1}$, when $\tau \ll \beta$).  
Moreover, this procedure requires a rigorous bound, which will be derived in Section \ref{sec:Bounding2pt}. 

\begin{figure*}[t]
\centering
   \includegraphics[width=0.6\textwidth]{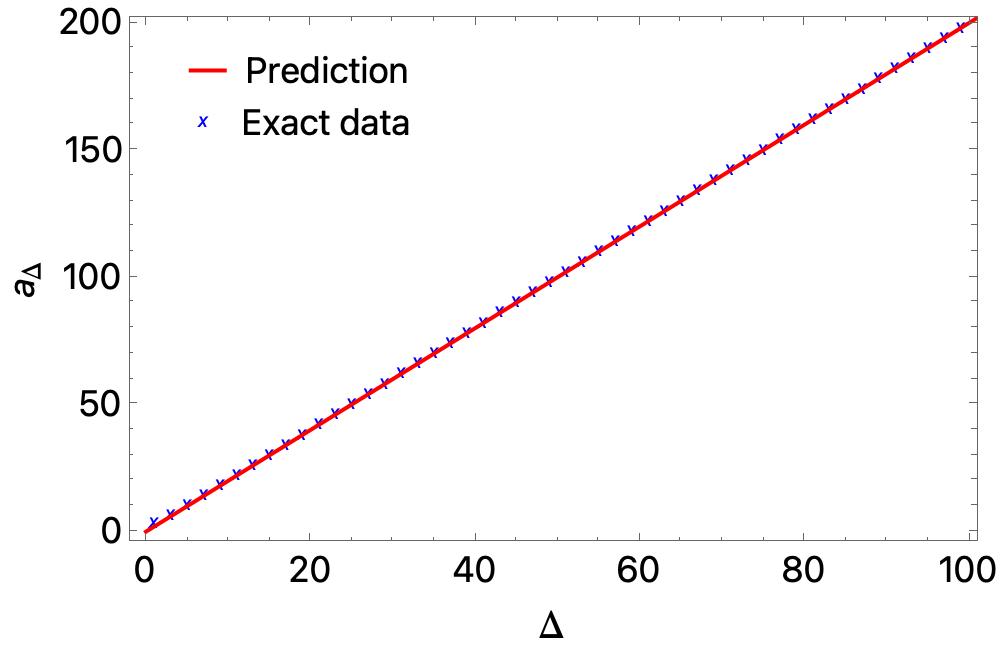}
  \caption{\emph{Exact OPE coefficients $a_{\Delta}$ in the case of a free scalar theory in four dimensions, or a GFF in one dimension with $\Delta_\phi = 1$. Each one is represented by a blue dot. The plot should be compared with the Tauberian prediction, represented as a continuous red line.} }\label{figtauberiancheck}
\end{figure*}

\begin{figure}[h!]
  \centering
\begin{subfigure}[t]{.45\textwidth}
\hspace{0.2cm}
  \includegraphics[width=1.008\textwidth]{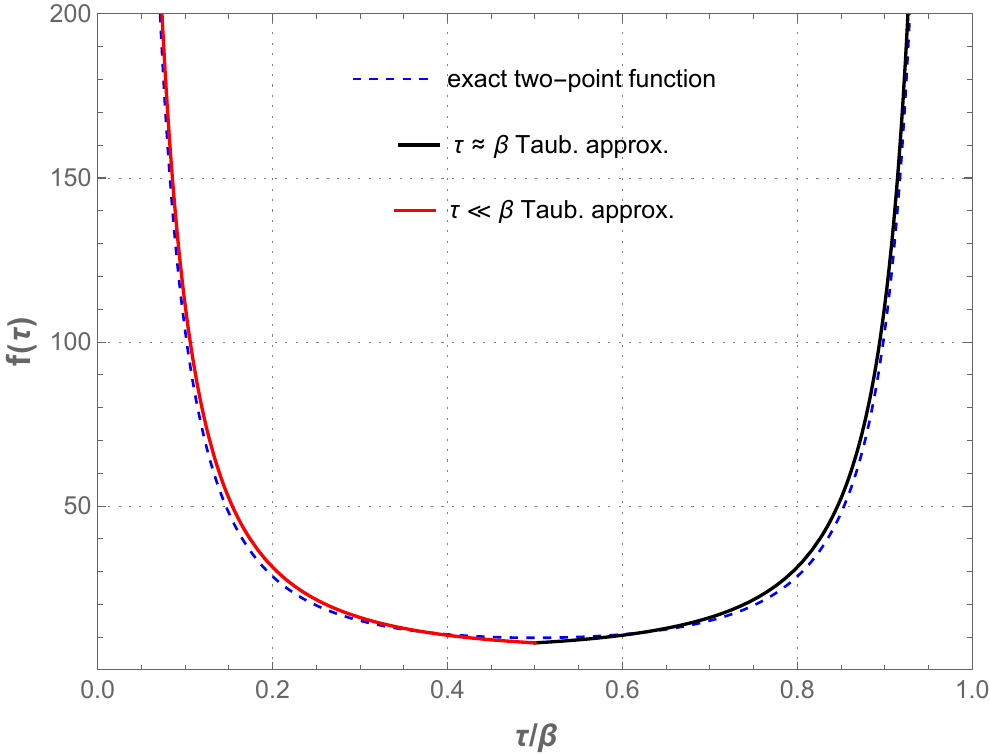}
\end{subfigure}%
\hfill
\begin{subfigure}[t]{.45\textwidth}
  \centering
  \includegraphics[width=\linewidth]{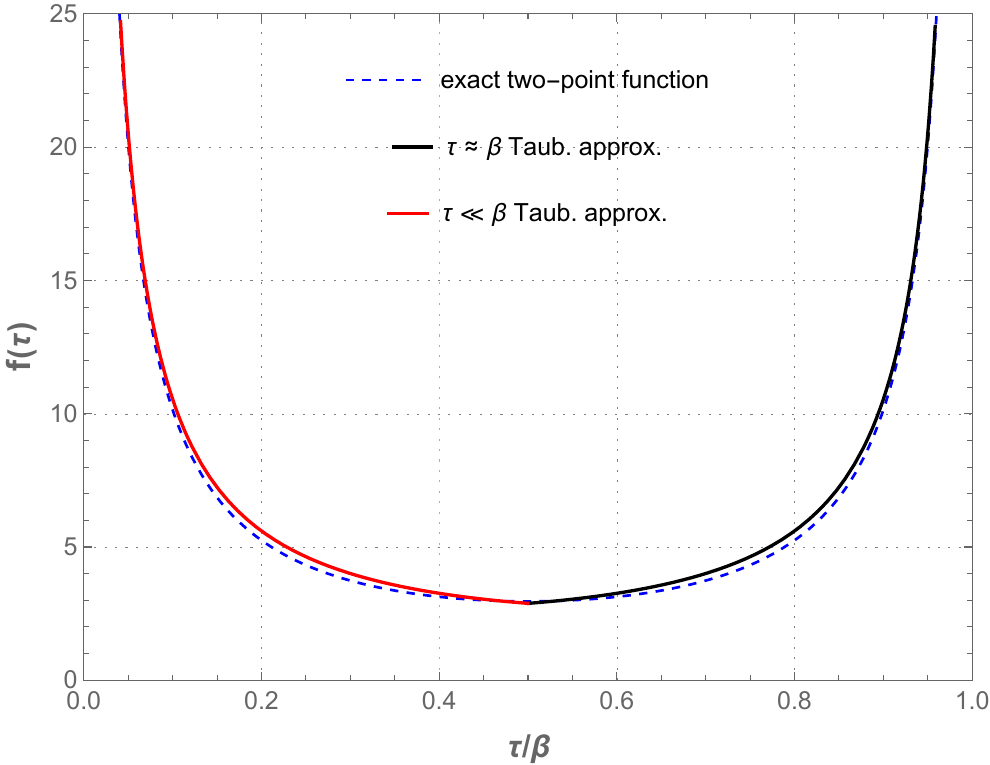}
\end{subfigure}%
\vfill
\begin{subfigure}[t]{.475\textwidth}
  \includegraphics[width=0.99\textwidth]{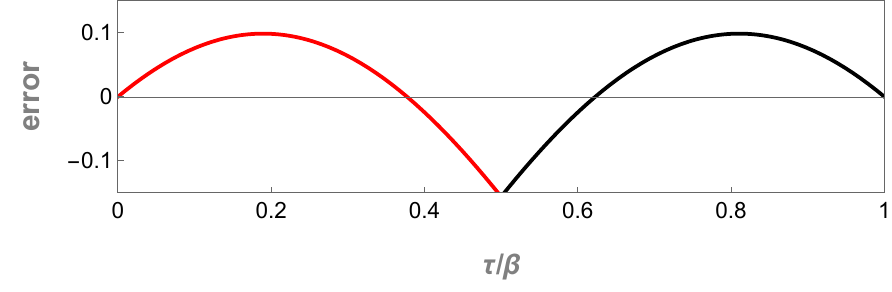}
  \caption{}
  \label{figtauberiantest1}
\end{subfigure}%
\hfill
\begin{subfigure}[t]{.475\textwidth}
  \centering
  \includegraphics[width=0.99\linewidth]{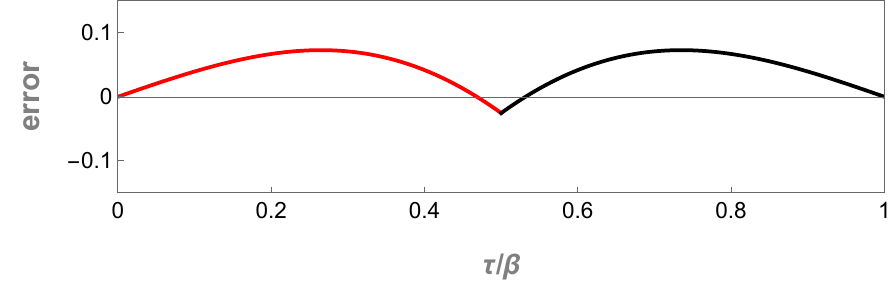}
  \caption{}
  \label{figtauberiantest2}
\end{subfigure}
\caption{\emph{Exact two-point function $\Braket{\phi(\tau)\phi(0)}_{\beta}$, represented as a dashed blue line, and its Tauberian approximation, divided into an approximation from left (continuous red line), and an approximation from right (continuous black line). We also present the relative difference between the exact two-point function and the Tauberian estimation, i.e. the quantity $(\langle \phi(\tau)\phi(0)\rangle_\beta^\text{Taub.}-\langle \phi(\tau)\phi(0)\rangle_\beta^\text{exact.})/\langle \phi(\tau)\phi(0)\rangle_\beta^\text{exact.}$. \textbf{Left panel (a)} The case of \underline{four dimensional free theory}. The maximum of the absolute value of the relative error is, as expected, in $\tau/\beta = 1/2$ and it is $\sim 0.156$. \textbf{Right panel (b)} The case of the \underline{$\mathrm O(N)$ model at large $N$}. The maximum of the absolute value of the relative error is, as expected, in $\tau/\beta \sim 0.25$ and it is $\sim 0.07$. See Section \ref{eq:O(N)largeN} for more details.}}\label{figtauberianTest}
\end{figure}

\noindent The result \eqref{eq:2ptfunctionapprox} can be tested for thermal two-point functions that can be computed exactly: in Fig. \ref{figtauberianTest}, we present the comparison with the two-point function \eqref{D=1GFF}, that describes both GFF with $\Delta_{\phi}=1$ and a four-dimensional fundamental free scalar. In the same Fig. we also plot the same comparison for the $\mathrm O(N)$ model at large $N$.
In four dimensions the operators appearing in the OPE have dimensions \begin{equation}
    \Delta_J = 2 + J \ , \hspace{1 cm} \Delta_{\mathds{1}} = 0 \ ,
\end{equation}
and are therefore all integers. 
Since the exact two-point function is available, it is easy to extract all the OPE coefficients $a_{\Delta}$ appearing in the equation \eqref{eq:TwoPointFunctionInt}. Thus, we can compute the exact partial sums of such coefficients, which correspond to partial integrations of the spectral density \eqref{eq:densitydef} and the OPE coefficients at large dimensions: the latter quantity can be compared with Tauberian prediction, as presented in Fig. \ref{figtauberiancheck}, where we found perfect agreement. In Fig. \ref{figtauberianTest} instead the exact two-point function and its Tauberian approximation were plotted together, showing very good agreement in the neighborhoods of the divergences of the two-point function. What is more it seems from the plot in Fig. \ref{figtauberianTest} that the approximation is also quite stable when $\tau \to \beta/2$. As we will discuss in the next session the first correction to the Tauberian theorems is given by a relative term of the form $1/\log(\Delta)$. This first correction translates in a correction of order $\log(\tau/\beta)$ or $\log((\beta-\tau)/\beta)$ (see next Section) for the $\tau \sim \beta$ and the $\tau/\beta \sim 0$ approximations respectively. The coefficient in front of this term is theory dependent, however the log correction itself contributes with a relative error of at most $\log 2 \sim  0.6$ (for the case of the free theory), which could explain the good agreement even in the region where we do not expect it.

\subsubsection{Bounding thermal two-point functions for $r=0$}\label{sec:Bounding2pt}
In this Section, we will discuss the possibility of employing the Tauberian theorem to derive bounds on the thermal two-point functions of two identical scalar operators when restricted to zero spatial coordinates.  
Let us start by considering the OPE expansion \eqref{eq:TwoPointFunctionInt} in the regime $\tau \sim \beta$, where the heavy operators dominate the sum. As shown by the equation \eqref{eq:largeDcoeff}, in this limit the OPE coefficients can be estimated by using the Tauberian theorem
\begin{equation}
    \langle \phi(\tau) \phi(0)\rangle_\beta \overset{\tau/ \beta \sim 1}{\sim} \sum_{\Delta \gg 1} a_{\Delta} \frac{\tau^{\Delta-2\Delta_\phi}}{\beta^{\Delta}} = \sum_{\Delta \gg 1} \left(\frac{\Delta^{2\Delta_{\phi}-1}}{\Gamma(2\Delta_\phi)} \delta \Delta \right) \frac{\tau^{\Delta-2\Delta_\phi}}{\beta^{\Delta}} \ .
\end{equation}  
It is possible to impose a bound on $\delta \Delta$ (whose definition can be found in Section \ref{ssec: heavy}). The reasoning is the following: among all the possible heavy operators appearing in the OPE, we surely have \emph{double-twist operators} $[\phi \phi]_{p,q}$. The distance $\delta \Delta$ between two double-twist operators can be estimated by noticing that their conformal dimensions can be written as
\begin{equation}
    \Delta_{\text{double twist}}  = 2 \Delta_\phi + J + \frac{\#}{J}+ \frac{\#}{J^2}+\ldots \ ,
\end{equation} 
which implies that $\delta \Delta \le  \delta J+ \mathcal{O}(1/J)$ and therefore $\delta \Delta\le 2$, since the one-point functions of odd spin operators are set to be zero by symmetry arguments \cite{Iliesiu:2018fao}. The inequality symbol ensures the possibility of having other operators in between the two double-twist operators. We conclude that the existence of the double-twist operators class in the heavy sector of the $\phi \times \phi$ OPE sets an upper bound on $\delta \Delta$, which can easily be translated into an upper bound for the two-point function 
\begin{equation} \label{eq: sumapp}
    \langle \phi(\tau) \phi(0)\rangle_\beta \overset{\tau/ \beta \sim 1}{\sim}  \sum_{\Delta \gg 1} \left(\frac{\Delta^{2\Delta_{\phi}-1}}{\Gamma(2\Delta_\phi)} \delta \Delta \right) \frac{\tau^{\Delta-2\Delta_\phi}}{\beta^{\Delta}} \leq 2 \sum_{\Delta \gg 1} \left(\frac{\Delta^{2\Delta_{\phi}-1}}{\Gamma(2\Delta_\phi)}  \right) \frac{\tau^{\Delta-2\Delta_\phi}}{\beta^{\Delta}}  \ .
\end{equation}  
The right-hand side can be evaluated by replacing the sum with an integration: this introduces an additional upper bound and we get
\begin{equation}
    \langle \phi(\tau) \phi(0)\rangle_\beta \overset{\tau/ \beta \sim 1}{\lesssim} 2\left[\tau \log\left(\frac{\tau}{\beta}\right)\right]^{-2\Delta_\phi} \ .
\end{equation}
This bound is correct only for $\tau \to \beta$ or, by using the KMS dual equation, 
\begin{equation}
    \langle \phi(\tau) \phi(0)\rangle_\beta \overset{\tau/ \beta \ll 1}{\lesssim} 2\left[(\beta-\tau) \log\left(\frac{\beta-\tau}{\beta}\right)\right]^{-2\Delta_\phi}  \ .
\end{equation}
The bound gets less precise as $\tau$ gets closer to the value $\beta / 2 $. This can be verified by computing the first correction to the Tauberian approximation. This is given by an inverse logarithmic correction (see Chapter VII of \cite{tauberian} and equation (4.13) of \cite{Pappadopulo:2012jk})
\begin{equation}
    \int_0^\Delta \rho(\widetilde \Delta) \de \widetilde \Delta \overset{\Delta \to \infty}{ \sim} \frac{\Delta^{2\Delta_\phi}}{\Gamma(2\Delta_\phi+1)} \left[1+\mathcal{O}\left(\frac{1}{\log \Delta}\right)\right] \ .
\end{equation}
This implies that the bounds derived in this Section are correct up to terms of order $\mathcal{O}(\log(\tau/\beta))$ and $\mathcal{O}(\log(1-\tau/\beta))$ for $\tau/\beta \sim 1$ and $\tau/\beta \ll 1$ respectively, as shown explicitly in Appendix \ref{appendix:DetailsOn2ptbounds}. Implementing this correction, we conclude that the two-point function is bounded as follows
\begin{equation}
    \langle \phi(\tau) \phi(0)\rangle \lesssim \left\{\begin{matrix}
 \displaystyle  2\left[1 + \mathcal{O}\left(\log\left(\frac{\tau}{\beta}\right) \right)\right]\left[\tau \log\left(\frac{\tau}{\beta}\right)\right]^{-2\Delta_\phi} & \tau/\beta \sim 1 \\ 
 \\ 
  \displaystyle 2\left[1 + \mathcal{O}\left(\log \left(\frac{\beta-\tau}{\beta} \right)\right)\right]\left[(\beta-\tau) \log\left(\frac{\beta-\tau}{\beta}\right)\right]^{-2\Delta_\phi}  & \tau/\beta \ll 1
\end{matrix}\right. \ .
\end{equation}
These bounds imply that, as expected, the point of the thermal circle corresponding to the less precise bound is $\tau = \beta/2$, since the correction is more relevant in such a point. This is reminiscent of the zero temperature case \cite{Kraus:2018pax}. In the zero temperature case, the four-point function is approximated by the contribution from light states; also in that case channel duality plays a central role and in the self-dual point, the error is maximal.
\newline To discuss a possible bound from below we have to invoke \emph{unitarity}. Unitarity at zero temperature is important to have reflection positivity, ensuring the positivity of the Euclidean two-point function. We could just cite the axiomatic approach \cite{Frohlich:1975pf}, but we justify here why this is true by considering the \emph{thermofield double}. The latter is a formalism in which the finite temperature Hilbert space $\mathcal{H}_{\text{TFD}}$ is constructed using two copies of the zero temperature Hilbert space $\mathcal{H}$
\begin{equation}
    \mathcal H_\text{TFD} = \mathcal H \otimes \mathcal H \ ,
\end{equation} 
and the thermal vacuum state is defined as 
\begin{equation}
    \ket{0}_\beta = \sum_{\mathcal E} e^{-\frac{\beta}{2} \mathcal E} \ket{\mathcal E}\otimes \ket{\mathcal E} \ . 
\end{equation}
 Physical operators $\mathcal O$ can now be promoted to finite temperature operators by simply tensoring them with the identity, namely $\mathcal O \otimes \op 1$. The crucial point is that we can use reflection positivity of the Hilbert space $\mathcal H_\text{TFD}$, or simply the fact that for every state in the Hilbert space $\langle \psi | \psi \rangle \ge 0$. In this way, we have a bound from below for the thermal two-point functions 
 \begin{equation}
     \langle \phi(\tau) \phi(0)\rangle_\beta \ge 0 \ . 
 \end{equation}
\subsubsection{Tauberian theorem and error for the truncated sum rules}\label{sec:error}
In Section \ref{sec:KMSTCFT} we used the KMS condition and parity to predict thermal one-point function coefficients of light operators, introducing the hypothesis of a gap between the light and the heavy sectors. Given this hypothesis, we assumed that the contribution coming from heavy operators in \eqref{eq:BootstrapFor1D} was negligible (this was naively justified by looking at the examples in Section \ref{sec:KMSTCFT}). In this Section, we use the Tauberian theorem to quantify the validity of this assumption. 

\begin{figure*}[t]
\centering
   \includegraphics[width=0.55\textwidth]{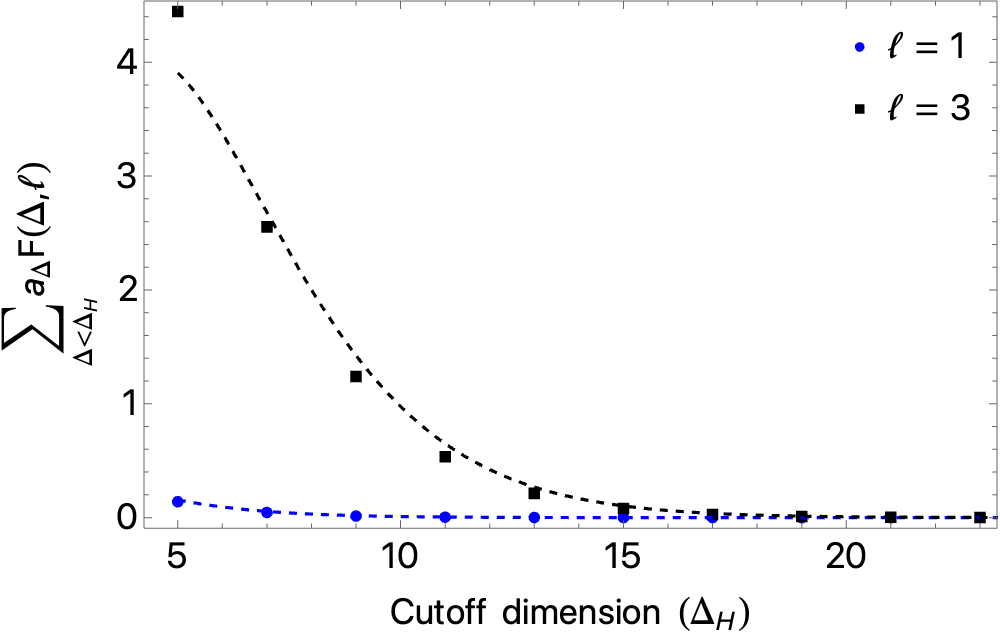}
  \caption{\emph{The (absolute value) of the difference between left and right side of equation \eqref{eq:BootstrapFor1D} for different values of $\ell$ for the two point function of fundamental scalars in the 3d $\mathrm O(N)$ model at large $N$. The explicit solution of the two-point function at zero spatial coordinate is given in Section \ref{eq:O(N)largeN}: the operators contributing in the sum rules are powers of the Hubbard-Stratonovich field $\sigma^m$ and the double twists of the fundamental field $[\phi_i\phi_i]_{n,l}$. The result is compared to the prediction from the Tauberian theorem given in \eqref{eq: gapppsum} (dashed lines): the numerical result are in agreement with the expectation.}}\label{figtauberiancheck2}
\end{figure*}

\noindent Let us consider a theory with $N$ light operators,  separated by a gap from an infinite tower of heavy operators. Let us assume that the first heavy operator has conformal dimension $\Delta_H \gg 1$: the sum rule \eqref{eq:BootstrapFor1D} can be written as 
\begin{equation}
    \sum_{\Delta <\Delta_H} a_{\Delta} F(\Delta,2p-1)+\sum_{\Delta \ge \Delta_H} a_{\Delta} F(\Delta,2p-1) = 0 \ , \quad p=1, \dots, N \ , \label{eq: gapsum}
\end{equation}
where $F(\Delta,2p-1)$ is defined in equation \eqref{eq:largegapped eq}. By employing the Tauberian theorem we can now justify the fact that the sum over heavy operators with dimensions $\Delta \ge \Delta_H$ can be neglected: plugging the result \eqref{eq:largeDcoeff} in the sum gives
\begin{equation}
  \sum_{\Delta \ge \Delta_H} a_{\Delta} F(\Delta,2p-1) \sim \sum_{\Delta \ge \Delta_H} \frac{\Delta^{2\Delta_\phi-1}}{\Gamma(2\Delta_\phi)} \delta\Delta F(\Delta,2p-1) \ .  \label{eq: sumheav}
\end{equation}
By assuming that $\Delta_H \ge 2\Delta_\phi+2N-2$, the sum \eqref{eq: sumheav} is made only by positive terms. Moreover, in Section \ref{sec:Bounding2pt} we argued that it is possible to impose the upper bound\footnote{Let us recall that this is true since $\delta \Delta=2$ is the minimal distance between two double-twist operators.} $\delta \Delta \le 2$. Therefore the sum \eqref{eq: sumheav} is bounded 
\begin{equation}
  \sum_{\Delta \ge \Delta_H} a_{\Delta} F(\Delta,2p-1) \lesssim 2  \sum_{\Delta \ge \Delta_H}\frac{\Delta^{2\Delta_\phi-1}}{\Gamma(2\Delta_\phi)} F(\Delta,2p-1) \ .  \label{eq: ineq}
\end{equation}
The right-hand side of the inequality \eqref{eq: ineq} can be estimated by turning the sum into an integration
\begin{equation}\label{eq:predictionerr}
  2 \int_{\Delta_H}^\infty \de \Delta \frac{\Delta^{2\Delta_\phi-1}}{\Gamma(2\Delta_\phi)} F(\Delta,2p-1)\overset{\Delta_H \gg 1}{\sim} \frac{1}{2^{\Delta_H}} \frac{\Delta_H^{2(\Delta_\phi+p-1)}}{\Gamma(2\Delta_\phi)\log 2} \ .
\end{equation}
We conclude that the sum rule \eqref{eq: gapsum} can be rewritten by bounding the sum over the heavy operators as follows
\begin{equation}
    \sum_{\Delta <\Delta_H} a_{\Delta} F(\Delta,2p-1)=\mathcal{O}\left( \frac{\Delta_H^{2(\Delta_\phi+p-1)}}{2^{\Delta_H}}\right)  \ , \quad p=1, \dots, N \ . \label{eq: gapppsum}
\end{equation}
The behaviour above can be checked for theories which admit an analytical expression for the two-point functions. In particular in Figs. \ref{fig:D1GFF}, \ref{fig:D3GFF}, \ref{fig:Ising2dcheck}, \ref{fig:YL2dCheck} and \ref{figtauberiancheck1} we show the behaviour of the right hand side of equation \eqref{eq:BootstrapFor1D} as a function of the cutoff dimension and we compared it with the contribution of the identity, represented by the left hand side of equation \eqref{eq:BootstrapFor1D}, for $2d$ CFTs, generalized free theories and the $\mathrm O(N)$ model at large $N$. It is possible to add the identity contribution to the right hand side of \eqref{eq:BootstrapFor1D} and compare directly with \eqref{eq: gapppsum}. We performed this analysis and we compare the result to \eqref{eq: gapppsum} finding agreement: as an example we show in Fig. \ref{figtauberiancheck2} the case of the $3d$ $\mathrm O(N)$ model at large $N$.
The interpretation is the following: if we consider a large gap theory with $N$ light operators, we can determine their thermal one-point function coefficients as shown in Section \ref{ssec: larggap}, making an approximation whose error is quantified by the right-hand side of the equation \eqref{eq: gapppsum}. Notice that the error increases with the number of light operators $N$ that we want to determine, and decreases exponentially with $\Delta_{H}$. We observe that the smaller $p$ is, the smaller the error introduced in the equation \eqref{eq: gapppsum}. To be concrete and to give quantitative predictions on the order of magnitude of a realistic error, let us consider the following OPE
\begin{equation}
    \phi(\tau) \times \phi(0)  \sim  \mathds{1} + \sum_i \mathcal O_{i}+\text{heavy operators} \ .
\end{equation}
The operators ($\mathds{1},\{ \mathcal{O}_{i \neq 0}\}$) belong to the light sector, whereas the first operator in the heavy sector has conformal dimension chosen to be $\Delta_H = 12$. Thus the equation \eqref{eq: gapppsum} takes the form for $p=1$
\begin{equation} \label{eq: cons}
    2\Delta_\phi+ \mathcal{O} \left(\frac{12^{2\Delta_\phi}}{2^{12}}\right)=\sum_{i} \frac{a_{\Delta_i}}{2^{\Delta_i}} (\Delta_i-2\Delta_\phi) \ .
\end{equation}
In the case of $\Delta_\phi = 1$, the equation \eqref{eq: cons} relates $a_{\Delta_i}$ with an error of order $4 \%$ on the equation. We stress that this error does not depend on the number of light operators, but only on the dimension of the external field $\phi$ (in the case above $\Delta_\phi = 1$), the dimension used as a cutoff $\Delta_H$ (in this case $\Delta_H \sim 12$) and the number $p$ identifying the equation we are solving in the list of sum rules \eqref{eq:BootstrapFor1D}. In order to solve the truncated system the number of equations has to correspond to the number of relevant fields (excluding the identity). However, it is important to notice that the number of light operators $N$ can potentially be a monotonic increasing function of $\Delta_{H}$. This fact can prevent a good estimation of all the light operators' one-point function coefficients $b_{\mathcal{O}_L}$ since the errors, increasing with $p = N$, quickly become very large. Even when the estimation of the error does not allow solving for the light operators' one-point function, it is still possible to use the equation \eqref{eq: gapppsum} with a small value of $p$ to connect OPE coefficients associated with different light operators.

\subsubsection{An example: three-dimensional $\mathrm O(N)$ model at large $N$}\label{eq:O(N)largeN}
Up to this Section, our results have been checked on (generalized) free theories or two-dimensional models. In this Section, we want to test our results on an interacting, higher-dimensional theory: the three-dimensional $\mathrm O(N)$ model at large $N$. This theory, whose central charge is $c = N c_\text{free}$ to leading order in $N$, has been studied in the literature both at zero temperature (see e.g. \cite{Sachdev:1993pr, Chubukov:1993aau, Kos:2013tga, Kos:2015mba, Kos:2016ysd,Dey:2016zbg,Alday:2019clp} and citation thereof) and at finite temperature \cite{Petkou:2018ynm, Chai:2020zgq, Chai:2020onq, David:2023uya}. We start by reviewing a few features of this model. The Lagrangian of the theory can be written after applying the Hubbard-Stratanovich transformation to the $(\phi_i^2)^2$ coupling, introducing the Hubbard-Stratonovich field $\sigma$ \begin{equation}\label{eq:ONLagrangian}
    \mathcal L = \frac{1}{2}(\partial \phi_i)^2+ \frac{1}{2}\sigma \phi_i^2- \frac{\sigma^2}{4 \lambda} \ ,
\end{equation} 
and the critical point is defined for $\lambda\to \infty$. The thermal mass of this theory\footnote{In a CFT at finite temperature, the thermal mass $m_{\text{th}}$ is the coefficient ruling the exponential decay of the thermal two-point function: $\Braket{\phi(\tau, r) \phi(0,0)}_{\beta}\overset{r \to \infty}{\longrightarrow} \Braket{\phi}_{\beta}^{2}+\mathcal{O}\left(e^{-m_{\text{th}}r} \right) \ .$}  is given by the thermal one-point function of $\sigma$, which can be computed as in \cite{Sachdev:1992py} (see Appendix C of \cite{Iliesiu:2018fao} for a review)\begin{equation}
    \langle \sigma\rangle_\beta = m_\text{th}^2 = \frac{4}{\beta^2} \log^2 \left(\frac{1+\sqrt 5}{2}\right) \ .
\end{equation} 
The thermal two-point function was also studied in \cite{Iliesiu:2018fao} and can be obtained by inverse Fourier transforming the propagator of the field $\phi$: \begin{equation}
    \langle \phi_i(\tau,r) \phi_j(0,0)\rangle_\beta = \delta_{ij} \sum_{m = -\infty}^\infty \int \frac{\de^2 k}{(2\pi)^2} \frac{e^{-i \vec k \cdot \vec x-i \omega_m \tau}}{\omega_n^2+\vec k^2+m_\text{th}^2} = \delta_{ij} \sum_{m = -\infty}^\infty \frac{e^{-m_\text{th}\sqrt{(\tau+m\beta)^2+r^2}}}{\sqrt{(\tau+m\beta)^2+r^2}}  \ . \label{eq: 2pton}
\end{equation}
In the spirit of this work, we reduce the two-point function \eqref{eq: 2pton} to zero spatial coordinates, so that we have 
\begin{multline}\label{eq:ON2pt}
     \langle \phi_i(\tau) \phi_j(0)\rangle_\beta  =\delta_{ij} \sum_{m = -\infty}^\infty \frac{e^{-m_\text{th} |\tau+m\beta|}}{|\tau+m\beta|}  = \\ = \delta_{ij} \frac{e^{m_\text{th} (\tau -\beta)}}{\beta}\Phi \left(e^{-m_\text{th}\beta},1,\frac{\beta-\tau}{\beta} \right)+\delta_{ij}\frac{e^{-m_\text{th} \tau }}{\beta} \Phi \left(e^{-m_\text{th}\beta},1,\frac{\tau}{\beta} \right) \ ,
\end{multline}
for $\tau \in (0,\beta)$\footnote{This breaks explicitly parity: $\tau \to -\tau$. However it is easy to recover it by imposing it explicitly.}, where $\Phi$ is the Lerch transcendent.\footnote{In our notation the Lerch transcendent is defined as \begin{equation}
    \Phi(z,s,\alpha) = \sum_{n = 0}^\infty \frac{z^n}{(n+\alpha)^s} \ .
\end{equation}This function is sometimes called Hurwitz–Lerch $\phi$-function.} In the specific case of $\Delta = 1/2$, the Lerch transcendent enjoys an hypergeometric representation. In the specific case of the $\mathrm O(N)$ model at large $N$ the two point function is 
\begin{equation}\label{eq:O(N)largeNtwoPointFunction}
     \langle \phi_i(\tau) \phi_j(0)\rangle_\beta = \delta_{ij} \frac{e^{-m_\text{th}(\beta-\tau)}}{\beta-\tau}  \, {}_{2}F_{1}{\left[\left.\genfrac..{0pt}{}{1,\frac{\beta-\tau}{\beta}}{1+\frac{\beta-\tau}{\beta}}\right| e^{-m_{\text{th}}\beta}\right]}+ \delta_{ij}\frac{e^{-m_\text{th}\tau}}{\tau} \, {}_{2}F_{1}{\left[\left.\genfrac..{0pt}{}{1,\frac{\tau}{\beta}}{1+\frac{\tau}{\beta}}\right| e^{-m_{\text{th}}\beta}\right]}  \ .
\end{equation}
The conformal dimensions of the fundamental fields are
\begin{equation} \label{eq: confon}
    \Delta_{\phi_i} = \frac{1}{2}+\mathcal{O}\left(\frac{1}{N}\right) \ , \hspace{1 cm}  \Delta_{\sigma} = 2+\mathcal{O}\left(\frac{1}{N}\right) \ ,
\end{equation}
therefore we expect only operators with integer conformal dimensions in the $\phi_i \times \phi_j$ OPE since operators constructed out of an odd number of copies of $\phi_i$ are forbidden by $\mathbb Z_2$ symmetry. This implies that the OPE expansion of the two-point function \eqref{eq:O(N)largeNtwoPointFunction} is equivalent to its Taylor expansion. The coefficients $a_\Delta$ can be extracted and compared with the prediction coming from the Tauberian theorem. In particular, recalling the equation \eqref{eq:TauberianMath} and considering the conformal dimensions \eqref{eq: confon} we expect the partial sums over the OPE coefficients to satisfy the Tauberian prediction
\begin{equation}
    \int_0^{\Delta} \rho(\widetilde \Delta) \de \widetilde \Delta \overset{\Delta \to \infty}{\sim}  \Delta \ . \label{eq: taub1}
\end{equation}
We can also formulate a Tauberian prediction for the OPE coefficients of heavy operators: employing the equation \eqref{eq:largeDcoeff} for the $\mathrm O(N)$ model at large $N$, i.e. fixing $\Delta_\phi = 1/2$ and $\delta \Delta= 2$ we have   
\begin{equation}
    a_\Delta  \overset{\Delta \to \infty}{\sim} 2 \ . \label{eq: taub2}
\end{equation}
\begin{figure*}[t]
\centering
\begin{subfigure}[t]{.49\textwidth}
   \includegraphics[width=\textwidth]{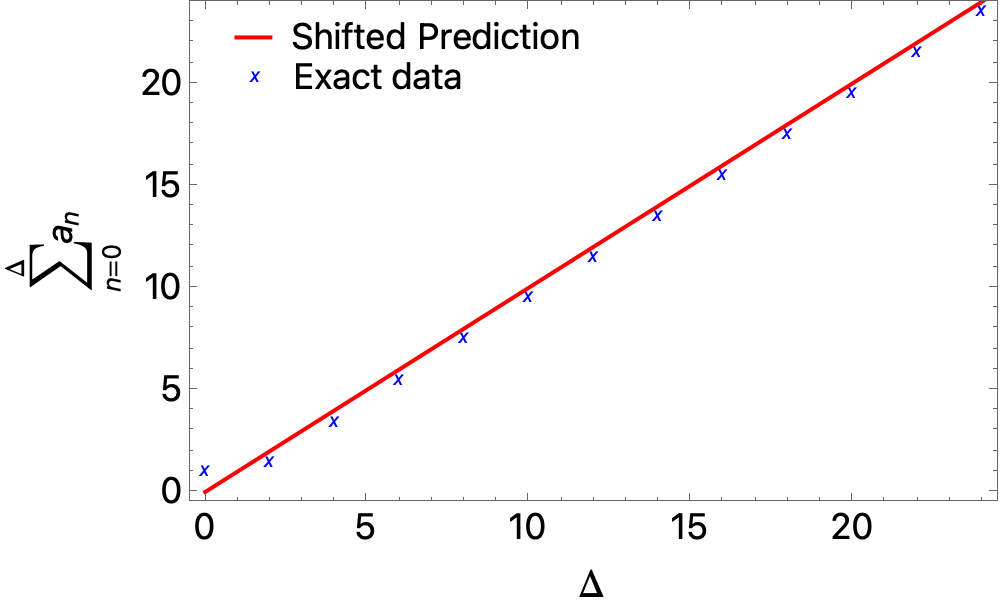}
  \caption{}\label{fig:ONtau}
\end{subfigure}%
\hfill
\begin{subfigure}[t]{.49\textwidth}
    \includegraphics[width=\textwidth]{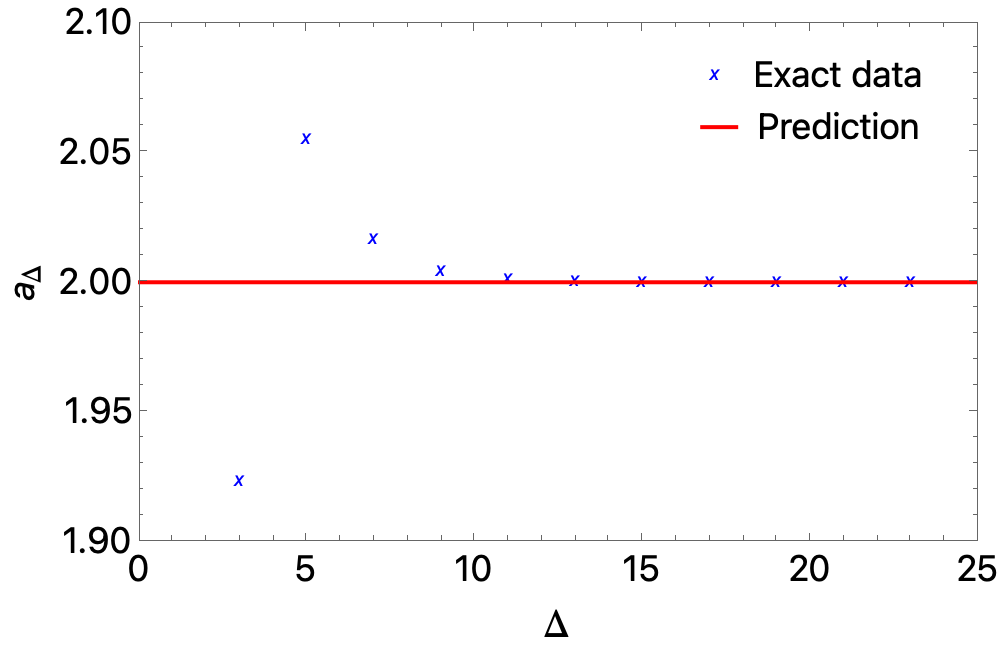}
    \caption{}\label{fig:ONan}
\end{subfigure}
  \caption{\emph{ In the \textbf{left panel (a)}, we plot the behavior of the partial sums over the exact OPE coefficients (indicated by blue dots) together with the Tauberian prediction (indicated by a straight red line). Since the Tauberian theorem only predicts the leading behavior of the sum, it returns a linear behavior in $\Delta$. To properly compare the exact data with the prediction, the latter needs to be vertically shifted by a constant $k=2$: this is not predicted by the theorem, which can only return the slope of the linear behavior. In the \textbf{right panel (b)}, we report the comparison between the exact OPE coefficients (represented by blue dots) and the Tauberian prediction \eqref{eq: taub2} (represented by a constant, straight red line). Notice that this plot is in agreement with the depiction in Fig. \ref{fig:ExpectedBev}, since the sign of the heavy operators OPE coefficients stabilizes.}}
  \label{Fig:tauON}
\end{figure*}
Both predictions can be compared with the OPE coefficients extracted directly from the Taylor expansion of the two-point function \eqref{eq:O(N)largeNtwoPointFunction}, as shown in Fig. \ref{Fig:tauON}: in Fig. \ref{fig:ONtau} the partial sums over the exact OPE coefficients are compared with the Tauberian prediction \eqref{eq: taub1} while in Fig. \ref{fig:ONan} the OPE coefficients are directly compared with the Tauberian prediction \eqref{eq: taub2}.
\subsection{A comment on unequal external dimensions}
The procedure we employed in Section \ref{ssec: heur} can be generalized for the case of unequal external dimensions. Let us consider the OPE 
\begin{equation}
    \mathcal O_1(\tau) \times \mathcal O_2(0) \sim f_{\mathcal O_1 \mathcal O_2 \mathcal O_3} \tau^{\alpha} \mathcal O_3(\tau) + \ldots \ ,
\end{equation}
where $\alpha = \Delta_{\mathcal O_3}-\Delta_{\mathcal O_2}-\Delta_{\mathcal O_1}$ and $\mathcal O_3$ is the lightest operator in the OPE.
Karamata's argument presented in Appendix \ref{appendix:OPEdensityProof} can be adopted for any $\alpha \in \mathbb R$. However when $\alpha >-1$ the boundedness of the OPE coefficients is not a sufficient condition: in this case, we really need the positivity of all the OPE coefficients. Under this assumption we know that \begin{equation}
    \langle \mathcal O_1(\tau) \mathcal O_2(0) \rangle_\beta \overset{\tau \to \beta}{\sim} a_{\mathcal O_3}(\beta-\tau)^{\alpha}+\ldots \ .
\end{equation}
Therefore we can use the Tauberian theorem in this scenario as well. Similarly to the equal external dimension case, we introduce a density of states $\rho(\Delta)$
\begin{equation}
    \langle \mathcal O_1(\tau) \mathcal O_2(0) \rangle_\beta = \int_0^\infty  \de \Delta \ \rho(\Delta) \tau^{\Delta-\Delta_2-\Delta_1} \ .
\end{equation}
Following the heuristic argument given in Section \ref{ssec: deriv} or the proof in Appendix \ref{appendix:Derivation}, we find 
\begin{equation}
    \int_0^{\Delta}   \rho(\widetilde\Delta) \de \widetilde \Delta \overset{\Delta \to \infty}{\sim} \frac{ a_{\mathcal O_3}}{\Gamma(\alpha+1)} \frac{1}{\Delta^{\alpha}} \ , \hspace{1 cm} \Delta \gg 1  \ .
\end{equation}
As already observed, in the case in which $\alpha >-1$ the positivity of the OPE coefficients is necessary since the boundedness from below is not sufficient. Indeed by subtracting the OPE coefficients with its negative minimum value, we produce in the integral a term which we expect to diverge as $(1-\tau)^{-1}$, which could dominate over the physical leading term. Nevertheless, we showed in this work that there are simple theories as (generalized) free theories or two-dimensional models in which OPE coefficients are always positive: for these theories, it is possible to study unequal external dimensions as well.
The simplest application concerns the two-point function between the stress-energy tensor $T$ and a primary operator $\mathcal O$. In this case, the leading term in the OPE, i.e. the lightest operator, is $\mathcal O$ itself \cite{Osborn:1993cr}. This means that if the operator $\mathcal O$ has a non-zero one-point function, then the OPE density for heavy operators is given by \begin{equation}
    \rho_{T \mathcal O}(\Delta) \overset{\Delta \to \infty}{\sim} \frac{a_{\mathcal O }}{\Gamma(d)} \frac{1}{\Delta^{d-1}} \ .
\end{equation}

\section{Discussion}\label{eq: discussion}
In this work, we studied
finite temperature CFTs. Leveraging the OPE, parity invariance, and the KMS condition, we derived explicit sum rules for one-point functions. These sum rules, as explicitly presented in equation \eqref{eq:sumrule}, provide an estimation for one-point functions of light operators.

While equation \eqref{eq:sumrule} encodes valuable information, the lack of positivity in one-point functions precludes a straightforward numerical exploration. Nevertheless, we do not rule out the possibility that more sophisticated analytical and numerical techniques could extract meaningful physical data from these equations. Furthermore, we demonstrated how, in theories with a large gap, when the infinite sum in \eqref{eq:sumrule} can be truncated, the consistency conditions of the finite temperature theory also constrain the zero temperature CFT: we show it for the conformal dimensions. It remains unclear whether these constraints are equivalent to zero temperature constraints (e.g. the crossing equation) or represent additional constraints on the CFT data. We look forward to updating on progress in these directions in future work.

The majority of the paper is dedicated to analyzing the case where the spatial coordinates of a thermal two-point function are set to zero. While this simplifies the correlation function significantly, it renders the OPE insensitive to the spin of the various operators involved. Consequently, OPE coefficients in this regime become weighted sums of the standard OPE coefficients of operators sharing the same conformal dimension. Nevertheless, we establish that when these quantities correspond to heavy operators, they all share the same sign, indicating that the OPE coefficients are bounded. This assumption is crucial for utilizing Tauberian theorems to estimate the asymptotic behavior of the OPE density, from which we extract the leading term of the OPE coefficients associated with heavy operators
\begin{equation}
    a_{\Delta} \overset{\Delta \to \infty}{\sim}\frac{\Delta^{2\Delta_\phi-1}}{\Gamma(2\Delta_\phi)}\delta \Delta \ ,
\end{equation}
where $\delta \Delta$ is the minimal distance between the conformal dimension $\Delta$ and the next one appearing in the OPE. The Tauberian theorems, similar to the zero-temperature case \cite{Qiao:2017xif}, establish a connection between light operators and heavy operators in the other KMS channel.

Furthermore, the two-point functions at zero spatial coordinates can be bounded by utilizing the Tauberian estimation of the OPE density with errors on this bound under control. Several checks are provided for (generalized) free theories, two-dimensional models, and the three-dimensional $\mathrm O(N)$ model at large $N$, where the two-point functions and, consequently, the OPE coefficients are analytically under control. We also address scenarios where the external operators do not share the same conformal dimensions, demonstrating how the Tauberian estimation could be employed to approximate the values of the tail of heavy operators in the sum rules outlined in equation \eqref{eq:sumrule}. However the boundedness of the coefficients can not be proved for those cases.

Continuing this exploration, it would be highly intriguing to further investigate sum rules as in equations \eqref{eq:sumrule} and high-energy behaviors. More sophisticated techniques with respect to the usual numerical bootstrap are needed in order to solve the sum rules. In this way of thinking the Tauberian theorem derived in this work could help in re-summing the tale of heavy operators (in analogy with the zero temperature case \cite{Su:2022xnj}), avoiding a naive truncation of the OPE.
Another recent approach we believe is worth exploring is the reinforcement-learning techniques recently developed in \cite{Kantor:2021kbx,Kantor:2021jpz,Niarchos:2023lot}.
In order to do that it is important to understand how the behavior predicted by the theorem changes when the spin is reintroduced in equation \eqref{eq:BootstrapFor1D}, i.e. when we consider the correlator at non-zero spatial coordinates and the degeneracy in spins is splitted. In addition,  it was noticed recently \cite{Petkou:2021zhg,Karydas:2023ufs} that thermal one-point functions of free theories are related to Feynman graphs corresponding to certain Fishnet models. It would be interesting to understand if there is a relation between the sum rules in \eqref{eq:sumrule} and the Tauberian approximations and this class of graphs. 
What is more, yet another promising direction regards the \emph{embedding space} and the \emph{ambient space} formalism,  proposed in \cite{Parisini:2022wkb,Parisini:2023nbd}: it would be interesting to compare such results with the one presented in this work. An implicit version of the Cardy formula was also proposed in \cite{El-Showk:2011yvt} and interpreted as a consequence of the conformal broken Ward identities at finite temperature in \cite{Marchetto:2023fcw}. Since the Cardy formula in two dimensions can be derived by modular invariance it is natural to ask if in higher dimension there is a generalization of modularity (see \cite{El-Showk:2011yvt,Marchetto:2023fcw,Allameh:2024qqp} for recent developments): it would interesting to explore more in this direction and possibly connect it with Tauberian theorems.

Notably, finite temperature CFTs have holographic dualities with black holes in AdS, and, recent, very interesting, applications in this context have been discussed (see, for example, \cite{Alday:2020eua,Dodelson:2023vrw, Dodelson:2023nnr,Caron-Huot:2022lff} for $d \ge 3$). It is particularly interesting to understand the implications of sum rules and Tauberian approximations in holographic theory to make predictions on the gravity dual side. It was pointed out that two-point functions in thermal CFTs with an holographic dual can be written in terms of the quasi-normal modes of the black hole in AdS \cite{Dodelson:2023vrw}: it would be extremely interesting to understand if there is a direct connection between OPE coefficients of heavy operators and those quantities. In order to answer this question one should study the bootstrap problem in momentum space \cite{Manenti:2019wxs}. We are working on these topics and hope to present our progress in the near future.

\acknowledgments
    It is a pleasure to thank Ant\'onio Antunes, Julien Barrat, Apratim Kaviraj, Diego Rodriguez-Gomez, Jorge Russo, Slava Rychkov, and Sasha Zhiboedov for very useful discussions. We thank Julien Barrat and  Apratim Kaviraj for very useful comments on the draft. We especially thank Sridip Pal for many useful discussions on Tauberian theorems which impelled us to further detail our explanation and to add Appendix \ref{app:TaubOPE} to our original manuscript.  EP is supported by ERC-2021-CoG - BrokenSymmetries 101044226. AM and EP have benefited from the German Research Foundation DFG under Germany’s Excellence Strategy – EXC 2121 Quantum Universe – 390833306. The authors would like to express special thanks to the Mainz Institute for Theoretical Physics (MITP) of the Cluster of Excellence PRISMA$^+$ (project ID 39083149), for its hospitality and support. AM is grateful to the DESY theory workshop 2023 \textit{New Perspectives in Conformal Field Theory and Gravity} where some of the results of this paper were presented.
    
\appendix 

\section{Derivation of the equation \eqref{eq:sumrule}}\label{appendix:Derivation}
    In this Appendix, we derive the sum rules in equation \eqref{eq:sumrule}. The starting point is the OPE \eqref{eq:OPE} and the general consistency relation \eqref{eq:GeneralDContraint} encoding the KMS condition.
    As a first step, it is convenient to expand the Gegenbauer polynomials as
    \begin{equation} \label{eq: gegdef}
       C_{J}^{(\nu)}\left(\frac{\tau}{\sqrt{\tau^2+r^2}} \right) = \sum_{k = 0}^{[J/2]}2^{J-2k}  (-1)^k \frac{\Gamma\left(J-k+\nu\right)}{\Gamma(\nu)k! (J-2k)!} \left(\frac{\tau}{\sqrt{r^2+\tau^2}}\right)^{J-2k} \ .
   \end{equation}
   Plugging the decomposition  in the equation \eqref{eq:OPE} \begin{equation}
        f(\tau,r)  =  \sum_{\mathcal O \in \phi \times \phi}\sum_{k = 0}^{[J/2]}\frac{a_{\mathcal O}}{\beta^{\Delta}} (-1)^k \frac{\Gamma\left(J-k+\nu\right)}{\Gamma(\nu)k! (J-2k)!}2^{J-2k}  |\tau^2+r^2|^{\frac{h-2 \Delta_{\phi}+2k}{2}} \tau^{J-2k} \ ,
   \end{equation}
   where we adopted the twist variable \eqref{eq: twist}. By using the (generalized) binomial theorem, we obtain
  \begin{equation}
       f(\tau,r)  =   \sum_{\mathcal O \in \phi \times \phi}\sum_{k = 0}^{[J/2]}\sum_{n = 0}^\infty \frac{a_{\mathcal O}}{\beta^{\Delta}} (-1)^k \frac{\Gamma\left(J-k+\nu\right)}{\Gamma(\nu)k! (J-2k)!}2^{J-2k}  \binom{\frac{h-2\Delta_{\phi}+2k}{2}}{n}\tau^{\Delta-2\Delta_\phi-2 n} r^{2 n} \ . \label{eq: r=0}
  \end{equation}
  We now compute the function in $\frac{\beta}{2} \pm \tau$, and we apply the binomial theorem  a second time  
  \begin{multline}
      f\left(\frac{\beta}{2} \pm \tau,r\right)  = \sum_{\mathcal O \in \phi \times \phi}\sum_{k = 0}^{[J/2]}\sum_{n, \ell = 0}^\infty \frac{a_{\mathcal O}}{\beta^{2\Delta_{\phi}+2n+\ell}} (-1)^k \frac{\Gamma\left(J-k+\nu\right)}{\Gamma(\nu)k! (J-2k)!}2^{-h-2k+2 \Delta_{\phi}+2n +\ell}\times \\ \times \binom{\Delta-2\Delta_\phi-2n}{\ell}  \binom{\frac{h-2\Delta_{\phi}+2k}{2}}{n} \left(\pm\tau\right)^{\ell} r^{2 n} \ .
  \end{multline}
  It is crucial to notice that the expression is sensitive to the sign difference in $\frac{\beta}{2} \pm \tau$ if and only if $\ell$ is an odd positive integer. If we impose the KMS condition \eqref{eq:GeneralDContraint}, we get 
 \begin{multline}
      \sum_{\mathcal O \in \phi \times \phi}\sum_{k = 0}^{[J/2]}\sum_{n = 0}^\infty \frac{a_{\mathcal O}}{\beta^{2n}} (-1)^k \frac{\Gamma\left(J-k+\nu\right)}{\Gamma(\nu)k! (J-2k)!}2^{-h-2k+2n}\times \\ \times \binom{\Delta-2\Delta_\phi-2n}{\ell}  \binom{\frac{h-2\Delta_{\phi}+2k}{2}}{n} r^{2 n}  = 0 \ ,
  \end{multline}
  for a fixed $\ell \in 2 \mathbb N +1$. These equations must be true for any $n\in \mathbb N$ and for any $\ell \in 2 \mathbb N +1$, therefore they reduce to
  \begin{equation}
      \sum_{\mathcal O \in \phi \times \phi}\sum_{k = 0}^{[J/2]}a_{\mathcal O} (-1)^k \frac{\Gamma\left(J-k+\nu\right)}{\Gamma(\nu)k! (J-2k)!}2^{-h-2k}   \binom{\Delta-2\Delta_\phi-2n}{\ell}  \binom{\frac{h-2\Delta_{\phi}+2k}{2}}{n}  = 0 \ , \label{eq: sumrulesappe}
  \end{equation}
  Once we re-sum over $k$, the equation can be recast as   \begin{equation}
      \sum_{\mathcal O \in \phi \times \phi} b_{\mathcal O} f_{\mathcal O \phi \phi} F_{\ell, n}(h,J) = 0 \ ,
  \end{equation}
  where we used the definition \eqref{eq: def a} of the $a_{\co}$ coefficients (the normalization constants $c_{\mathcal{O}}$ have been set equal to 1 identically)
  \begin{equation}
      a_{\mathcal O}= \frac{f_{\phi\phi\mathcal O}b_{\mathcal O}}{c_{\mathcal O}} \frac{J!}{2^J (\nu)_J} \ , 
  \end{equation}
  and we introduced the function
  \begin{equation} \label{eq: vector}
      F_{\ell, n}(h,J)=\frac{1}{2^{h+J}} \binom{\frac{h-2 \Delta_{\phi}}{2}}{n} \binom{h+J-2 \Delta_{\phi} -2 n }{\ell} \, {}_{3}F_{2}{\left[\left.\genfrac..{0pt}{}{\frac{1-J}{2},-\frac{J}{2}, \frac{h}{2}-\Delta_{\phi} +1}{\frac{h}{2}-\Delta_{\phi}-n +1,-J-\nu +1}\right| 1\right]} \ .
  \end{equation}
  This concludes the derivation of the sum rules \eqref{eq:sumrule}.
  \paragraph{Reduction to $r=0$}
   Now, we want to specialize the sum rules to the case of two operators lying on the same thermal circle. This limit is achieved by setting $r=0$. By looking at  equation \eqref{eq: r=0}, we see that the only significative equation is returned by setting $n=0$. This considerably simplifies the expression \eqref{eq: vector}
   \begin{equation} 
      F_{\ell, 0}(h,J)=\frac{1}{2^{h+J}}  \binom{h+J-2 \Delta_{\phi} }{\ell} \, {}_{2}F_{1}{\left[\left.\genfrac..{0pt}{}{\frac{1-J}{2},-\frac{J}{2}}{-J-\nu +1}\right| 1\right]} \ .
  \end{equation}
   This expression of $F_{\ell,0}$ is such that the sum rule \eqref{eq:sumrule} is equivalent, in the $n =0$ reduction, to \eqref{eq:BootstrapFor1D}. To see this we use the fact that, by symmetry arguments, $J \in 2 \mathbb N$, and $\nu \in \mathbb N /2$ by definition: hence, it can be shown that \begin{equation}
      C_{J}^{(\nu)}(1)=\frac{2^J (\nu)_{J}}{J!} \, {}_{2}F_{1}{\left[\left.\genfrac..{0pt}{}{\frac{1-J}{2},-\frac{J}{2}}{-J-\nu +1}\right| 1\right]}\ .
  \end{equation}
 The sum rules \eqref{eq: sumrulesappe} are now in the form, for a fixed odd integer $\ell$ \begin{equation}
     \sum_{\mathcal O \in \phi \times \phi} a_{\mathcal O} \, C_J^{(\nu)}(1) \frac{\Gamma(\Delta-2\Delta_\phi+1)}{2^\Delta \Gamma(\Delta-2\Delta_\phi-\ell+1)}=0   \ .
  \end{equation}
   Finally, since the only spin-dependent term is  $a_{\mathcal O} \, C_J^{(\nu)}(1)$, we can just define a weighted sum of thermal OPE coefficients 
   \begin{equation}
      a_{\Delta} \equiv \sum_{\mathcal{O} \in \phi \times \phi}^{\Delta \text{ fixed}}  a_{\mathcal O} \,  C_J^{(\nu)}(1) \ .
  \end{equation}
  Finally by separating the identity $\mathds{1}$ contribution \begin{equation}
      -\frac{\Gamma(2\Delta_\phi+k)}{\Gamma(2\Delta_\phi)}\ ,
  \end{equation} we end up with equation \eqref{eq:BootstrapFor1D}.
\section{Review of Tauberian theorems}\label{appendix:OPEdensityProof}
In this Appendix, we sketch the proof for the Tauberian theorem discussed in Section \ref{sec:tauberianOPE}: our proposal is an adaptation of Karamata's proof presented in Section 7.53 of \cite{tauberian2}.

\noindent Let us start by considering the integral \begin{equation}\label{eq:tauberianInteg}
    \int_0^\infty \de \Delta \ \rho(\Delta) x^{\Delta} \overset{x\to 1}{\sim} (1-x)^{-\alpha} \ , \quad \alpha>0  \ .
\end{equation}
The first step is to prove that for every polynomial $P(y)$ \begin{equation}
    \lim_{x \to 1} (1-x)^{\alpha} \int_0^\infty \de \Delta \ \rho(\Delta) x^\Delta P\left (x^\Delta\right) = \int_0^1 \de t \ P(t) \ .
\end{equation}
It is sufficient to prove the statement for $P(y) = y^n$. After a straightforward manipulation
\begin{equation}
      \lim_{x \to 1} (1-x)^{\alpha} \int_0^\infty \de \Delta \ \rho(\Delta) \left(x^{1+n}\right)^{\Delta}  =    \lim_{x \to 1} \left(\frac{1-x}{1-x^{n+1}}\right)^{\alpha }\left(1-x^{n+1}\right)^{\alpha}\int_0^\infty \de \Delta \ \rho(\Delta) \left(x^{1+n}\right)^{\Delta} \ .
\end{equation}
Recalling the hypothesis \eqref{eq:tauberianInteg}, the integral is diverging as 
\begin{equation}
    \int_0^\infty \de \Delta \ \rho(\Delta) \left(x^{1+n}\right)^{\Delta} \overset{x\to 1}{\sim} \left(1-x^{n+1}\right)^{-\alpha} \ ,
\end{equation}
and therefore 
\begin{equation}
     \lim_{x \to 1} (1-x)^{\alpha} \int_0^\infty \de \Delta \ \rho(\Delta) \left(x^{1+n}\right)^{\Delta} = \frac{1}{n+1} = \int_0^1 \de t \  t^{n}  \ ,
\end{equation}
as claimed above. In the following, let us consider a function $g(x)$ defined on the closed interval $[0,1]$. We want to approximate $g(x)$ with arbitrary precision from above and from below by using polynomials
\begin{equation}
    p(x) \le g(x) \le P(x) \ ,
\end{equation}
where 
\begin{equation}
   \int_0^1 \de x  \left (g(x)-p(x)\right)\le \epsilon \ , \hspace{1 cm}  \int_0^1 \de x \left (P(x)-g(x)\right)\le \epsilon \ ,
\end{equation}
for an arbitrarily small $\epsilon>0$. If $g(x)$ is a continuous function, it is sufficient to invoke Weierstrass's approximation theorem. However, we need this approximation also in the case of a function $g(x)$ that admits discontinuities of the first kind, i.e. for any point $c \in (0,1)$, both $\lim_{x \to c^{-}} g(x)$ and $\lim_{x \to c^{+}} g(x)$ exist, but $\lim_{x \to c^{-}} g(x) \neq \lim_{x \to c^{+}} g(x)$. The claim is correct also under this relaxed hypothesis: suppose without loss of generality that $\lim_{x \to c^{-}} g(x)< \lim_{x \to c^{+}} g(x)$, and let 
\begin{equation}
     \phi(x)= \left\{\begin{matrix}
 \displaystyle g(x)+\epsilon / 2  & \quad c-\delta > x \ \text{and} \ x>c \\ 
 \\ 
  \displaystyle \max\left\{l(x),g(x)+\epsilon/4\right\} & c-\delta \le x\le c
\end{matrix}\right. \ ,
\end{equation}
where $l(x)$ is a linear function defined in such a way that $l(c-\delta) = g(c-\delta)+\epsilon/2$ and $l(c) = \lim_{x \to c^{+}}g(x)+\epsilon/2$. Then $\phi(x)$ is continuous by construction and $\phi(x) > g(x)$: for small enough $\delta$, we can apply Weierstrass's approximation theorem to $\phi(x)$ and consequently polynomially bound $g(x)$ 
\begin{equation}
    P(x) \ge \phi(x) > g(x) \ .
\end{equation}
A similar construction can be repeated for the polynomial bound from below.
\newline The next step is to prove that for any $g(x)$ with only discontinuities of the first kind we have  \begin{equation}\label{eq1prooftauberian}
    \lim_{x \to 1} (1-x)^{\alpha} \int_0^\infty \de\Delta \ \rho(\Delta)  x^\Delta g\left(x^{\Delta}\right) = \int_0^1 \de t \ g(t) \ .
\end{equation}
Now, we need to introduce a crucial hypothesis: we consider the density $\rho(\Delta)$ to be positive definite
\begin{equation} \label{eq: taub hyp}
    \rho(\Delta) \ge 0 \ , \quad \text{for any } \Delta \ge 0 \ .
\end{equation}
Using this hypothesis, we can derive the following series of inequalities
\begin{equation}
\begin{split}
     \lim_{x \to 1} (1-x)^{\alpha} \int_0^\infty \de\Delta \  \rho(\Delta) x^\Delta g\left(x^{\Delta}\right) &\le  \lim_{x \to 1} (1-x)^{\alpha} \int_0^\infty \de\Delta \ \rho(\Delta) x^\Delta P\left(x^{\Delta}\right)   \\ &= \int_0^1 \de t \ P\left(t\right) \le  \int_0^1 \de t \ g\left(t\right)+\epsilon \ ,
\end{split}
\end{equation}
which leads to 
\begin{equation}
     \lim_{x \to 1} (1-x)^{\alpha} \int_0^\infty \de\Delta \ \rho(\Delta) x^\Delta g\left(x^{\Delta}\right) \le \int_0^1 \de t \ g\left(t\right) \ ,
\end{equation}
when $\epsilon \to 0$. Similarly, by using the polynomial bound from below, one can see that \begin{equation}
     \lim_{x \to 1} (1-x)^{\alpha} \int_0^\infty \de\Delta \ \rho(\Delta)  x^\Delta g\left(x^{\Delta}\right) \ge \int_0^1 \de t \ g\left(t\right) \ ,
\end{equation}
and therefore the equation \ref{eq1prooftauberian} is proved. 

\noindent The final step consists to choose a specific function $g(x)$ with a discontinuity of the first kind
\begin{equation}\label{eq:Karatag}
    g(x) = \left\{\begin{matrix}
0 \ , & 0 \le x \le e^{-1}\\ 
1/x \ , & e^{-1} < x \le 1 
\end{matrix}\right. \ .
\end{equation}
The function is engineered in such a way that
\begin{equation}
    \int_0^1 g(x) \de x = \int_{1/e}^1 \frac{\de x}{x} = 1 \ ,
\end{equation}
so that we can prove 
\begin{equation}
    \int_0^\infty \de \Delta \  \rho(\Delta) x^{\Delta} g\left (x^{\Delta}\right) \overset{x \to 1}{\sim} (1-x)^{-\alpha} \ .
\end{equation}
With this specific choice and parametrizing $x = e^{-1/\overline \Delta}$, we have 
\begin{equation}
    \int_0^{\overline \Delta}\de \Delta \ \rho(\Delta) =  \int_0^\infty \de \Delta \ \rho(\Delta) e^{-\Delta/\overline \Delta} g\left (e^{-\Delta/\overline \Delta}\right) \overset{\overline \Delta \to \infty}{\sim}\left(1-e^{-1/\overline \Delta}\right)^{-\alpha} \sim \overline \Delta^{\alpha } \ .
\end{equation}
The last equation is exactly the Tauberian theorem \eqref{eq:TauberianMath}.
\subsection{Tauberian theorems and boundedness}
In the scenario considered in this work, the hypothesis \eqref{eq: taub hyp} is not true in general. However, this hypothesis can be relaxed asking $\rho(\Delta)$ to be bounded from below, i.e. there exists a constant $C$ such that \begin{equation}
    \rho(\Delta) \ge - C \ .
\end{equation}
To see that this a sufficient condition for the proof in Appendix \ref{appendix:OPEdensityProof} to hold, it is enough to consider the same proof starting from the integral 
\begin{equation}\label{eqdivapp}
    \int_0^\infty \de \Delta \  \widetilde \rho(\Delta) x^\Delta \overset{x \to 1}{\sim} (1-x)^{-\alpha} \ , \quad \alpha>0 \ , 
\end{equation}
where $\widetilde \rho(\Delta) = \rho(\Delta)+C \ge 0$. If we isolate the contribution from the constant $C$, it generates a divergence of the type 
\begin{equation}
    C \int_0^\infty \de \Delta \ x^\Delta  \overset{x \to 1^-}{\sim} -\frac{C}{\log x} \ .
\end{equation}
In conclusion, if $\rho(\Delta)$ is bounded from below
\begin{equation}
    \int_0^\infty \de \Delta \  \rho(\Delta) x^\Delta \overset{x \to 1^-}{\sim} (1-x)^{-\alpha}+\frac{C}{\log(x)} \ .
\end{equation}
Observe that \begin{equation}
    \frac{C}{\log(x)}\sim \frac{C}{1-x}+ \text{regular terms} \ ,
\end{equation}
hence this divergence will be subleading with respect to the divergence in \eqref{eqdivapp} if and only if  $\alpha > 1$. In our scenario, this last condition gets translated into $\Delta_\phi > 1/2$.
\section{Derivation of the equation \eqref{eq:samesign}}\label{appendix:boundedness}
In this Appendix, we prove that the OPE coefficients $a_{\Delta}$, defined at zero spatial coordinates in equation \eqref{eq:twopintOPEZeror}, are bounded from below. To see this, we combine the KMS condition with the Euclidean inversion formula \cite{Iliesiu:2018fao}. In the following, $\beta$ will not play any role, so it will be set $\beta= 1$ without loss of generality. 

\noindent We start by packing the OPE coefficients into the function 
\begin{equation}
    a(\Delta)=\int_{0}^{1}\frac{\de \tau}{\tau}\tau^{2 \Delta_{\phi}-\Delta} f(\tau) \ , 
\end{equation}
where $f(\tau)=\Braket{\phi(\tau)\phi(0)}_{\beta}$ is the thermal two-point function at zero spatial coordinates.  The OPE coefficients $a_{\Delta}$ are encoded in the function $a(\Delta)$ as the residues of first-order poles
\begin{equation}
    a(\Delta) = \sum_{\widehat{\Delta}}  \frac{a_{\widehat \Delta}}{\widehat \Delta- \Delta} \ . \label{eq: adel}
\end{equation}

\noindent Then, we apply the KMS condition (plus parity) to the two-point function $f(\tau)$
\begin{equation}
    a(\Delta)=\int_{0}^{1}\frac{\de \tau}{\tau}\tau^{2 \Delta_{\phi}-\Delta} f(1-\tau)\ .
\end{equation}
The integrated can be expanded using the thermal OPE giving 
\begin{equation}
    a(\Delta)=\int_{0}^{1}\frac{\de \tau}{\tau}\tau^{2 \Delta_{\phi}-\Delta} \sum_{\widetilde{\Delta}}a_{\widetilde{\Delta}}(1-\tau)^{\widetilde{\Delta}-2 \Delta_{\phi}} \ .
\end{equation}
We can now switch integral and series, making sure to first regularize the integration to allow the operations to exchange 
\begin{equation}
    a(\Delta)=\lim_{\epsilon \rightarrow 0}\int_{\epsilon}^{1}\frac{\de \tau}{\tau}\tau^{2 \Delta_{\phi}-\Delta} \sum_{\widetilde{\Delta}}a_{\widetilde{\Delta}}(1-\tau)^{\widetilde{\Delta}-2 \Delta_{\phi}}=\lim_{\epsilon \rightarrow 0}\sum_{\widetilde{\Delta}}a_{\widetilde{\Delta}}\int_{\epsilon}^{1}\frac{\de \tau}{\tau}\tau^{2 \Delta_{\phi}-\Delta} (1-\tau)^{\widetilde{\Delta}-2 \Delta_{\phi}} \ .
\end{equation}
The integration can be explicitly performed
\begin{equation}
    a(\Delta)=-\lim_{\epsilon \rightarrow 0}\sum_{\widetilde{\Delta}}a_{\widetilde{\Delta}}\left[B_{\epsilon}\left(2 \Delta_{\phi}-\Delta, 1-2 \Delta_{\phi}+\widetilde{\Delta}\right)-B\left(2 \Delta_{\phi}-\Delta, 1-2 \Delta_{\phi}+\widetilde{\Delta} \right) \right] \ ,
\end{equation}
where the result is written in terms of the Euler Beta function
\begin{equation}
    B(a, b)=\frac{\Gamma(a) \Gamma(b)}{\Gamma(a+b)} \ .
\end{equation}
Furthermore, we make use of the identity  
\begin{equation}
    B_\epsilon(a,b)=B(a,b)-\frac{\epsilon^a (1-\epsilon)^b}{b}  {}_{2}F_{1}{\left[\left.\genfrac..{0pt}{}{1, a+b}{b+1}\right| 1-\epsilon\right]}
\end{equation}
to write the function $a(\Delta)$ as 
\begin{equation}
    a(\Delta)=\lim_{\epsilon \rightarrow 0}\sum_{\widetilde{\Delta}}a_{\widetilde{\Delta}}\frac{\epsilon^{2 \Delta_{\phi}-\Delta} (1-\epsilon)^{1-2 \Delta_{\phi}+\widetilde{\Delta}}}{1-2 \Delta_{\phi}+\widetilde{\Delta}}  {}_{2}F_{1}{\left[\left.\genfrac..{0pt}{}{1, \widetilde{\Delta}-\Delta+1}{2-2 \Delta_{\phi}+\widetilde{\Delta}}\right| 1-\epsilon\right]} \ .
\end{equation}
Having found an explicit expression of $a(\Delta)$, we need to compute its residues to extract a single coefficient $a_{\widehat \Delta}$ in terms of all the other OPE coefficients. We can easily project on $a_{\widehat{\Delta}}$ recalling the definition of $a(\Delta)$ in \eqref{eq: adel}
\begin{equation}
   a_{\widehat{\Delta}}=-\lim_{\Delta \rightarrow \widehat{\Delta}}\lim_{\epsilon \rightarrow 0}\sum_{\widetilde{\Delta}}a_{\widetilde{\Delta}} \left(\widehat{\Delta}-\Delta \right)\frac{\epsilon^{2 \Delta_{\phi}-\Delta} (1-\epsilon)^{1-2 \Delta_{\phi}+\widetilde{\Delta}}}{1-2 \Delta_{\phi}+\widetilde{\Delta}}  {}_{2}F_{1}{\left[\left.\genfrac..{0pt}{}{1, \widetilde{\Delta}-\Delta+1}{2-2 \Delta_{\phi}+\widetilde{\Delta}}\right| 1-\epsilon\right]} \ .
\end{equation}
To unclutter the notation, we can define $\Delta=\widehat{\Delta}+\eta$ and relabel $\widehat{\Delta} \rightarrow \Delta$
\begin{equation}
   a_{\Delta}=\lim_{\eta \rightarrow 0}\lim_{\epsilon \rightarrow 0}\sum_{\widetilde{\Delta}}a_{\widetilde{\Delta}}   \frac{\eta \epsilon^{2 \Delta_{\phi}-\Delta-\eta} (1-\epsilon)^{1-2 \Delta_{\phi}+\widetilde{\Delta}}}{1-2 \Delta_{\phi}+\widetilde{\Delta}}  {}_{2}F_{1}{\left[\left.\genfrac..{0pt}{}{1, \widetilde{\Delta}-\Delta-\eta+1}{2-2 \Delta_{\phi}+\widetilde{\Delta}}\right| 1-\epsilon\right]}\ . \label{eq: master}
\end{equation}
The equation \eqref{eq: master} is a consequence of the KMS condition and it expresses one single OPE coefficient in terms of all the other OPE coefficients. The expression is quite complicated and hard to use. However, we only need to prove that 
\begin{equation}
    a_{\Delta+\delta \Delta} = a_{\Delta}\left [1+ \mathcal{O} \left(\frac{1}{\Delta}\right)\right ] \ , \quad \Delta \gg 1 \ , 
\end{equation}
where it is understood that $\delta\Delta/\Delta\ll 1$. The idea is the following: we consider $a_{\Delta+\delta \Delta}$ 
\begin{equation}
   a_{\Delta+ \delta \Delta}=\lim_{\eta \rightarrow 0}\lim_{\epsilon \rightarrow 0}\sum_{\widetilde{\Delta}}a_{\widetilde{\Delta}}   \frac{\eta \epsilon^{2 \Delta_{\phi}-\Delta- \delta \Delta-\eta} (1-\epsilon)^{1-2 \Delta_{\phi}+\widetilde{\Delta}}}{1-2 \Delta_{\phi}+\widetilde{\Delta}}  {}_{2}F_{1}{\left[\left.\genfrac..{0pt}{}{1, \widetilde{\Delta}-\Delta- \delta \Delta-\eta+1}{2-2 \Delta_{\phi}+\widetilde{\Delta}}\right| 1-\epsilon\right]} \ , \label{eq: adeltadelta}
\end{equation}
and we expand the expression above in powers of $\delta \Delta$.  In general, $\delta \Delta$ cannot be considered a  small parameter: the best bound we have is $\delta \Delta \lesssim 2$ (see Section \ref{sec:tauberianOPE}). However, every order of the expansion in powers of $\delta \Delta$ is suppressed by a term of order $\mathcal{O}(1/\Delta)$. Let us consider the terms in the right-hand side of the equation \eqref{eq: adeltadelta} sensitive to $\delta \Delta$: the exponential of $\epsilon$, for $\Delta\gg 1$, can be approximated as
\begin{equation}
\epsilon^{2 \Delta_{\phi}-\Delta- \delta \Delta-\eta} \approx \epsilon^{2 \Delta_{\phi}-\Delta-\eta} \ , \quad \delta \Delta / \Delta \ll 1 \ , \Delta \gg 1 \ ,
\end{equation}
while the hypergeometric function is expanded as
\begin{multline} \label{eq: hyp der}
    {}_{2}F_{1}{\left[\left.\genfrac..{0pt}{}{1, \widetilde{\Delta}-\Delta- \delta \Delta-\eta+1}{2-2 \Delta_{\phi}+\widetilde{\Delta}}\right| 1-\epsilon\right]}={}_{2}F_{1}{\left[\left.\genfrac..{0pt}{}{1, \widetilde{\Delta}-\Delta-\eta+1}{2-2 \Delta_{\phi}+\widetilde{\Delta}}\right| 1-\epsilon\right]}+\\+\delta \Delta\sum_{n=0}^{\infty} \frac{(1)_{n+1}(\widetilde{\Delta}-\Delta-\eta+1)_{n+1}}{(2-2 \Delta_{\phi}+\widetilde{\Delta})_{n+1}}\frac{(1-\epsilon)^{n+1}}{(n+1)!}\sum_{m=0}^{n}\frac{1}{-m-\widetilde{\Delta}+\Delta+\eta-1}+\mathcal{O}(\delta \Delta^2) \ .
\end{multline}
It is clear that there are three different cases to study: \begin{enumerate}
    \item  $\Delta \gg 1$ and $\Delta \gg \widetilde \Delta$: in this case the first order term is suppressed by factors of order $\mathcal{O}\left(\frac{1}{\Delta}\right)$;
    \item  $\Delta \gg 1$ and $\Delta \ll \widetilde \Delta$: in this case the first order term is suppressed by factors of order $\mathcal{O}\left(\frac{1}{\widetilde \Delta}\right)$, which are consequently also of order $\mathcal{O}\left(\frac{1}{\Delta}\right)$;
    \item $\Delta \gg 1$ and $\Delta \sim \widetilde \Delta$: in this case, the sum over $m$ in the right-hand side of the equation \eqref{eq: hyp der} does not suppress the function. Instead, we must consider the term 
    \begin{equation}
        \frac{(\widetilde{\Delta}-\Delta-\eta+1)_{n+1}}{(2-2 \Delta_{\phi}+\widetilde{\Delta})_{n+1}} \sim \widetilde{\Delta}^{1-n} \sim \Delta^{1-n}  \ .
    \end{equation}
    Hence, the first order term is suppressed by factors of order $\mathcal{O}(1/\Delta)$. 
\end{enumerate}
The proof for higher orders in the expansion in powers of $\delta \Delta$ follows in the same way, such that the OPE coefficients for large $\Delta$ satisfy 
\begin{equation}
     a_{\Delta+\delta \Delta} = a_{\Delta}\left [1+ \mathcal{O} \left(\frac{1}{\Delta}\right)\right ] \ , \quad \Delta \gg 1 \ .
\end{equation}
\section{Details on the bound on two-point functions and its error}\label{appendix:DetailsOn2ptbounds}
In Section \ref{sec:tauberianOPE} we proved the existence of a bound for the thermal two-point function of two identical scalar operators, once it is reduced to zero spatial coordinates.  In this Appendix, we show some details of the computations. 

\noindent An important step in the derivation of the bound was the approximation of the right-hand side of the equation \eqref{eq: sumapp} with the integral
\begin{equation}
    2 \int_0^\infty \de \Delta  \frac{\Delta^{2\Delta_\phi-1}}{\Gamma(2\Delta_\phi)} \frac{\tau^{\Delta-2\Delta_\phi}}{\beta^{\Delta}} \ .
\end{equation}
The indefinite integration returns by definition \begin{equation}\label{eq:indefint1}
  2 \int \de \Delta \frac{\Delta^{2\Delta_\phi-1}}{\Gamma(2\Delta_\phi)} \frac{\tau^{\Delta-2\Delta_\phi}}{\beta^{\Delta}}=  -\frac{2 \Delta^{2\Delta_\phi}}{\tau^{2\Delta_\phi}} E_{1-2\Delta_\phi}\left (-\Delta \log\left(\frac{\tau}{\beta}\right)\right) \ ,
\end{equation}
where we introduced the \emph{exponential integral function} \begin{equation}
    E_n(z) = \int_1^\infty \frac{e^{-zt}}{t^n} \ \de t \ .
\end{equation}
By computing the indefinite integral \eqref{eq:indefint1} in the two extrema $\Delta=0$ and $\Delta = \infty$ we get \begin{equation}
    2 \int_0^\infty \de \Delta \frac{\Delta^{2\Delta_\phi-1}}{\Gamma(2\Delta_\phi)} \frac{\tau^{\Delta-2\Delta_\phi}}{\beta^{\Delta}}=2\left[\tau \log\left(\frac{\tau}{\beta}\right)\right]^{-2\Delta_\phi} \ .
\end{equation}
The KMS dual integration, i.e. when $\tau \to \beta-\tau$, can be performed in the same way.  Alternatively, we observe that we can just send $\tau \to \beta-\tau$ in the result since the integration is over $\Delta$.
\newline The integration to perform to address the logarithmic corrections to the Tauberian density is more complicated since we need to integrate 
\begin{equation}
      2\int_0^\infty \de \left[\frac{\Delta^{2\Delta_\phi}}{\Gamma(2\Delta_\phi+1)}\left(1+\frac{\kappa}{\log \Delta}\right)\right] \tau^{\Delta-2\Delta_\phi} \ , \label{eq: intkappa}
\end{equation}
where $\kappa$ is a constant. We can now use the fact that $\Delta \gg 1$ to justify replacing $\Delta$ with $\Delta+\epsilon$. This replacement has also the effect of regularizing the integral. We can say
\begin{equation}
    \frac{\Delta^{2\Delta_\phi}}{\Gamma(2\Delta_\phi+1)}\left(1+\frac{\kappa}{\log \Delta}\right) \overset{\Delta \to \infty}{\sim}  \frac{\Delta^{2\Delta_\phi}}{\Gamma(2\Delta_\phi+1)}\left(1+\sum_{k}\widetilde \kappa_k(\epsilon) \Delta^k\right)\  , \label{eq: kappa}
\end{equation}
where $\widetilde \kappa_k(\epsilon)$ contains $\kappa$ and the series coefficient of the $1/\log(\Delta+\epsilon)$ term; it is, in general, $\epsilon$ dependent. We fix $\epsilon$ so that the first term of the expansion is equivalent to the Tauberian leading term, ensuring the correct divergence of the integral. If we plug \eqref{eq: kappa} back into the integral \eqref{eq: intkappa}, we have a simpler expression to integrate and we can show that the leading term for $\tau/\beta \sim 1$ is  
\begin{equation}
   2\left[1 + \mathcal{O}\left(\log\left(\frac{\tau}{\beta}\right) \right)\right]\left[\tau \log\left(\frac{\tau}{\beta}\right)\right]^{-2\Delta_\phi}\ .
\end{equation}
\section{From Tauberian theorem to OPE coefficients}\label{app:TaubOPE}
We clarify here some mathematical details of the derivation of the OPE coefficients in equation \eqref{eq:largeDcoeff}. The subtraction procedure reported in Section \ref{ssec: heavy} is not mathematically precise, even if standard in physics literature.\footnote{The authors would like to thank Sridip Pal for very interesting discussions on this point.} In this Appendix we comment on the additional conditions to be added to rigorously derive equation \eqref{eq:largeDcoeff} from equation \eqref{eq:TauberianMath}. First, let us explain why the derivation presented in the main text is not completely rigorous. The key point is to recall that the Tauberian theorem, rigorously proven in this paper, is 
\begin{equation}\label{eq:taubreal}
    \int_0^\Delta \rho(\widetilde \Delta) \de \widetilde \Delta \overset{\Delta \to \infty}{\sim} \frac{\Delta^{2\Delta_\phi}}{\Gamma(2\Delta_\phi+1)} \left[1+\mathcal{O}\left(\frac{1}{\log\Delta}\right)\right]\ .
\end{equation}
When extracting the OPE coefficients following the prescription  \begin{equation}\label{eq:diff}
    a_{\Delta} = \left(\int_0^\Delta \de \widetilde \Delta -\int_0^{\Delta-\delta \Delta} \de \widetilde \Delta \right) \rho(\widetilde \Delta) \ ,
\end{equation}
the subtraction between the two leading terms gives \begin{equation}
   a_{\Delta} \overset{\Delta \to \infty}{\sim} \frac{\Delta^{2\Delta_\phi-1}}{\Gamma(2\Delta_\phi)} \delta \Delta \ .
\end{equation}
Possible problems could arise due to the error in the equation \eqref{eq:taubreal}. In the most general case, the error of the difference is 
\begin{equation}
    a_{\Delta} \overset{\Delta \to \infty}{\sim} \frac{\Delta^{2\Delta_\phi-1}}{\Gamma(2\Delta_\phi)} \delta \Delta\left[1+\mathcal{ O}\left(\frac{\Delta}{\log \Delta}\right)\right] \qquad \text{\emph{(real Taub. thm.)}} \ ,
\end{equation}
In this case, the prediction on $a_{\Delta}$ is spoiled since the error dominates over the leading term. Therefore the conclusion seems to be wrong. However, this estimation of the error is optimal by just using the tools of real analysis, but it was proved that going to the complex plane the estimation of the error can be improved to be of order $\mathcal{O}\left(\Delta^{2\Delta_\phi-1}\right)$. This result goes under the name of \emph{complex Tauberian theorem} \cite{complextauberian}. We will not show this result, since it is not the goal of this paper, but let us comment that even this improved estimation would not solve the problem since the error will compete with the leading behaviour
\begin{equation}
    a_{\Delta} \overset{\Delta \to \infty}{\sim} \frac{\Delta^{2\Delta_\phi-1}}{\Gamma(2\Delta_\phi)} \delta \Delta\Big[1+\mathcal{ O}\left(1\right)\Big] \qquad \text{\emph{(complex Taub. thm.)}} \ ,
\end{equation}
meaning that although we can prove that the OPE coefficients are of order $\mathcal{O}\left(\Delta^{2\Delta_\phi-1}\right)$, it is not possible in general to fix the numerical coefficients. 
\newline Nonetheless all the examples presented in this paper suggest that the result in equation \eqref{eq:largeDcoeff} is correct, even if not mathematically rigorous. Let us consider a two-dimensional primary-primary correlator with $\Delta = 2$ as an illustrative example: in this case, the OPE coefficients behave according to the prediction \eqref{eq:largeDcoeff}. 
\begin{figure*}[t]
\centering
\begin{subfigure}[t]{.49\textwidth}
   \includegraphics[width=\textwidth]{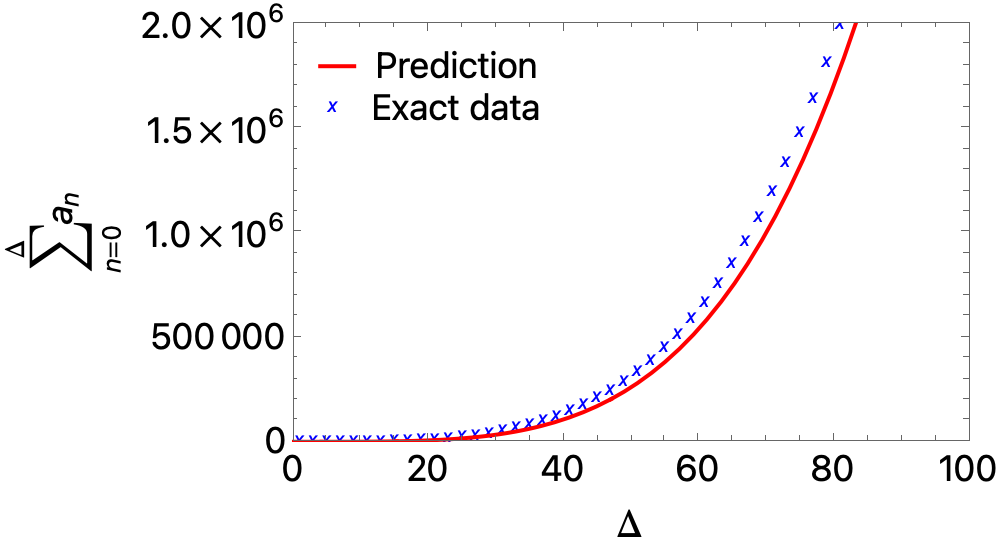}
  \caption{}\label{fig:Tauberro1}
\end{subfigure}%
\hfill
\begin{subfigure}[t]{.49\textwidth}
    \includegraphics[width=\textwidth]{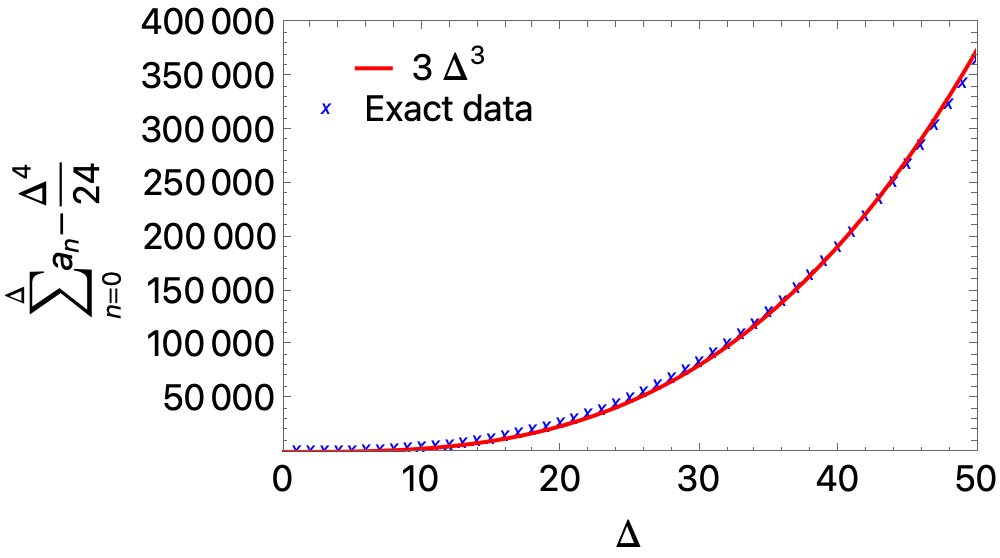}
    \caption{}\label{fig:Tauberro2}
\end{subfigure}
  \caption{\emph{ In the \textbf{left panel (a)}, we plot the behaviour of the partial sums over the exact OPE coefficients (indicated by blue dots) together with the Tauberian prediction (indicated by a straight red line). In the \textbf{right panel (b)}, we report the comparison between the computed error between the exact partial sums and the Tauberian prediction compared with a naive fit.}}
  \label{Fig:Tauberro}
\end{figure*}
\newline 

\noindent Let us start from the leading Tauberian: in Fig.  \ref{fig:Tauberro1} we propose a comparison between the leading Tauberian term, in this case simply $\frac{\Delta^4}{\Gamma(5)}$, and the exact partial sums. The Tauberian theorem, as expected, holds. More interesting is Fig. \ref{fig:Tauberro2} where we computed the difference between the exact partial sums and the Tauberian prediction $\frac{\Delta^4}{\Gamma(5)}$, reproducing the error on the leading behaviour. Interestingly enough, for heavy operators where the Tauberian theorem is supposed to hold, the error is cubic, specifically $\sim 3 \Delta^3$.\footnote{The details of the fit are not important for this argument since we are only interested in the behaviour and not in the exact coefficients of the fit.} We have now enough information to understand how the OPE coefficients of heavy operators behave. It is now possible to take the difference \eqref{eq:diff}:  since the subleading behaviour is fixed to be cubic, the error on the OPE coefficients can be computed and equation \eqref{eq:largeDcoeff} is correct for this example since the error only gives subleading terms. How can generalize this example? The key point is the following:  
if one can show that \begin{equation} \label{eq: strongTaub}
    \int_0^\Delta \rho(\widetilde \Delta) \ \de \widetilde \Delta \overset{\Delta \to \infty}{\sim}\frac{\Delta^{2\Delta_\phi}}{\Gamma(2\Delta_\phi+1)} \left(1+\frac{C}{\Delta}+\ldots\right) \ , \qquad C=\text{const.} \ ,
\end{equation}
then the subtraction \eqref{eq:diff} can be rigorously performed and the equation \eqref{eq:largeDcoeff} follows. The condition above can be considered as an \emph{analyticity} condition for the integral function. It would be interesting if this assumption could be formally relaxed to the analyticity of the OPE coefficients, which is a physically reasonable assumption.
\newline

\noindent Let us comment that the equation \eqref{eq: strongTaub} is stronger than the complex Tauberian generalization of equation \eqref{eq:taubreal}: in fact, the error is not only of order $\mathcal{O}\left(\Delta^{2\Delta_\phi-1}\right)$, but it can be written asymptotically as $C \Delta^{2\Delta_\phi-1}$. A simple example that satisfies the complex Tauberian theorem, but not its stronger version \eqref{eq: strongTaub}, could be a theory where 
\begin{equation} \label{eq: cosTaub}
    \int_0^\Delta \rho(\widetilde \Delta) \ \de \widetilde \Delta \overset{\Delta \to \infty}{\sim}\frac{\Delta^{2\Delta_\phi}}{\Gamma(2\Delta_\phi+1)} \left(1+\frac{\cos{\Delta}}{\Delta}+\ldots\right) \ .
\end{equation}
Although $\cos \Delta = \mathcal{O}(1)$, its oscillating behaviour makes the error important when we want to compute the OPE coefficients by performing the subtraction \eqref{eq:diff}. Let us remark that the scenario \eqref{eq: cosTaub} is just an illustrative example and we do not know if there is a physical theory (or even a function satisfying the hypothesis of the Tauberian theorem) that behaves like this.
\newline

\noindent In conclusion, to be completely rigorous in the derivation of the equation \eqref{eq:largeDcoeff} we have to assume that the leading term in the error of the Tauberian theorem is of the form \eqref{eq: strongTaub}. This can be checked to be the case for all the physical theories tested in this paper. If this condition can be proved from physical assumptions (considering the KMS condition, the Regge limit, etc.) is an open problem that we leave for the future. If proven, it would also be interesting to understand in different cases in which a Tauberian theorem of the type of equation \eqref{eq:TauberianMath} was proved, but a small size theorem was never been proved (see e.g. \cite{Qiao:2017xif}).

\newpage
	 \bibliographystyle{JHEP}
	 \bibliography{1db}
\end{document}